\documentclass[singlecolumn,amsmath, amssymb,floatfix]{revtex4}
\usepackage{hyperref}
\usepackage{epsfig}
\usepackage{epstopdf}
\pdfoutput=1
\usepackage{graphicx}
\usepackage{bm}
\usepackage{amssymb}
\usepackage{color}

\hypersetup{linktocpage}
\newcommand{\be}{\begin{equation}}
\newcommand{\ee}{\end{equation}}
\begin{document}
\title {Electron Glass Dynamics}

\author{Ariel Amir, Yuval Oreg, Yoseph Imry}

\affiliation { Department of Condensed Matter Physics, Weizmann
Institute of Science, Rehovot, 76100, Israel\\}
\begin {abstract}

Examples of glasses are abundant, yet it remains one of the phases
of matter whose understanding is very elusive . In recent years,
remarkable experiments have been performed on the dynamical aspects
of glasses. Electron glasses offer a particularly good example of
the 'trademarks' of glassy behavior, such as aging and slow
relaxations. In this work we review the experimental literature on
electron glasses, as well as the local mean-field theoretical framework put forward
in recent years to understand some of these results. We also present novel
theoretical results explaining the periodic aging experiment.

\end {abstract}


 \maketitle

\tableofcontents

\newpage
\section{Introduction}

Slow relaxations in nature have been observed long ago, in many
contexts. Two important examples are the 'stretched exponential'
behavior experimentally observed by Kohlrausch in the 19th century,
when studying discharge in a certain electrostatic setup known as a
Leiden jar \cite{Kohlrausch} and mechanical relaxations in silk
threads, used by Weber \cite{weber} to hang his magnets in his
seminal works on magnetism.

Slow relaxations are also important signatures of glassy behavior:
the Vogel-Fulcher law \cite{vogel} states that the timescale for the
thermalization of a glass diverges at a finite temperature. When
valid, this temperature is the phase-transition to the glassy phase,
below which non-ergodic behavior occurs: the system does not
equilibrate, and some states are never visited. Slow relaxations
have been known to occur in magnetic materials known as spin glasses
\cite{spinglass_aging1,spinglass_aging2,spinglass_aging3,spinglass_aging4},
vortices in superconductors in the glassy phase \cite{du}, and in
the electron glass system, to name but a few.

\begin{figure}[b!]
\includegraphics[width=0.5\textwidth]{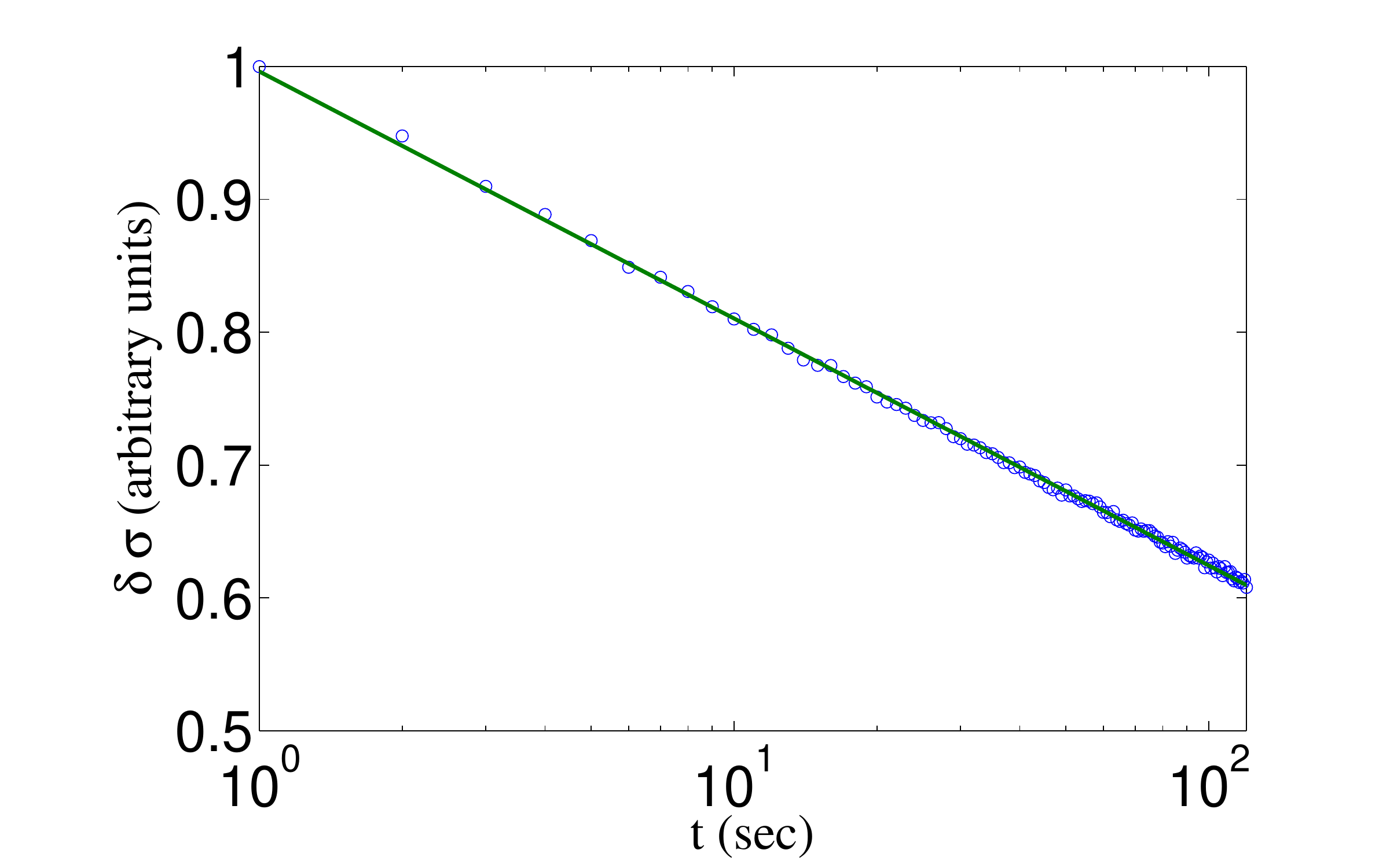}
\caption {Demonstration of a logarithmic relaxation in electron glasses, after a sudden change in gate voltage. In some cases, the logarithmic change in conductance can be measured from times of order of seconds to several days \cite{zvi_stress}. Data courtesy of Z. Ovadyahu.} \label
{log_plot}
\end{figure}

In many cases in nature the slow relaxations are logarithmic, over
several decades in time. These systems range from flux relaxation in
superconductors \cite{creep_super}, through crumpling paper \cite
{crumpling} where the volume is measured as a function of time, as
well as biological systems such as plant rheology\cite{plants}, and
electron glasses which we will focus on here. One should note that
it is not trivial to distinguish a logarithm from a stretched
exponential: one has to go to short time scales, where the power-law
behavior of the stretched exponential is distinctly different from
logarithmic. For this reason relaxations in the electron glass were
initially erroneously reported as stretched exponential. Fig.
\ref{log_plot} demonstrates a logarithmic relaxation after a sudden
change in the voltage of a gate coupled to the sample.

In this review we shall explain how one can understand these slow, logarithmic relaxations in electron glasses via the use of a local mean-field approximation, and also extend this to various aging protocols, in which the response
of the system to a perturbation depends also on the time the perturbation was applied for. We describe the experimental
details, but only briefly discuss theoretical frameworks other than the local mean-field approach.

The structure of the manuscript is as follows. We first explain different mechanism leading to a broad distribution of relaxation times, which will give rise to logarithmic relaxation over a broad range. We review the different experimental protocols demonstrating aging, and their results, focusing on granular aluminum and InO. We then go on to describe the local mean-field approach, and compare its prediction with the experimental results. We discuss the possible relation between $1/f$ noise and logarithmic relaxations. Next, more involved protocols related to memory effects in the system are reviewed, and the necessary theoretical framework for a qualitative understanding of them is given. We conclude with our view regarding the important, open questions in this field.

\section {Why a logarithm?}

One may decompose the logarithmic relaxation into a weighted sum of
decaying exponentials, by taking an inverse Laplace transform. This
gives a distribution of relaxation rate which is $P(\lambda) =
c/\lambda$, with $c=1/\rm{log}(\lambda_{\rm{max}}/\lambda_{\rm{min}})$, with
$\lambda_{\rm{max}}$ and $\lambda_{\rm{min}}$ the upper and lower cutoffs of
the distribution. Notice that the distribution of the logarithm of
$\lambda$ is uniform between the lower and upper cutoffs. Indeed,
taking its Laplace transform of $P(\lambda)$ yields, for times
$1/\lambda_{\rm{max}} \ll t \ll 1/\lambda_{\rm{min}}$ \cite{pollak2}:

\be \int e^{-\lambda t} P(\lambda) d\lambda \equiv L(t)=-\gamma_E-\rm{log}(\lambda_{\rm{min}}t) ,\ee with $\gamma_E$ the
Euler constant. Notice that the upper cutoff does not enter the
expression, but the lower one does.

In order to understand the logarithmic relaxations, it is sufficient to understand how a distribution $P(\lambda)= C/\lambda$ arises. This is the purpose
of the next section.

\subsection{Activated processes}

The simplest way to obtain a broad $1/\lambda$ distribution, is via
thermally activated processes: let us assume that a particle hops
between trapped states, and that the distribution of the energy
barriers between two adjacent traps is uniformly distributed in the
interval $[E_{\rm{min}},E_{\rm{max}}]$, $P(E) =
\frac{1}{E_{\rm{max}}-E_{\rm{min}}}$. If we assume thermally
activated processes, the rate $\lambda$ associated with a given
barrier is given by the Arrhenius law, $\lambda \sim e^{-E/T}$
(throughout the manuscript $k_B \equiv 1$). Calculating the
distribution of the rates, we find that $P(\lambda)d\lambda =
P(E)dE,$ and therefore $P(\lambda)\sim T/\lambda$.

In various experiments
\cite{zvi_aging1,zvi:TempProtocols,grenet_new} the temperature
dependence was shown to be much weaker, if at all measurable. For these cases, this
rules out the possibility of the above mechanism as being
responsible for the slow relaxations, and opts for a quantum
mechanism, nearly independent of temperature. Quantum tunneling is
the natural process to look at, in order to avoid the strong
dependence on temperature.

\subsection {Tunneling processes}
\label {tunnel}

 If we look at a particle hopping between localized states,
the tunneling rate is exponential in the distance $r$ (up to
polynomial corrections). If we assume that the spatial distribution
of localized states is uniform, we can readily calculate the
distribution of hopping rates between a given site and its
nearest-neighbor: the rate $\lambda \sim e^{-r/\xi}$, where $\xi$ is
the localization length, taken here as a constant for simplicity. We
have $P(\lambda)d\lambda = P(r)dr,$ with $P(r)$ the probability
distribution of having the nearest-neighbor at a distance $r$. It is
easier to calculate the cumulative of this distribution,
\emph{i.e.}, the probability $C(r)$ that all neighbors are more than
a distance $r$ away from a given site. The probability of a given
site to be more than a distance $r$ away is $p=1-V_d r^d/L^d$, with
$V_d=\pi^{d/2}/\Gamma(d/2+1)$ the volume of a $d$ dimensional unit
sphere and $L$ the system size. Thus, $C(r)=p^N$, where $N$ is the
number of sites. In the limit $N \gg 1$, this can be simplified to
give:

 \be P(r) = \frac{d
V_d}{\langle r \rangle} [r/\langle r \rangle]^{d-1}e^{-V_d
[r/\langle r \rangle]^d}, \ee $\langle r \rangle = \frac{L}{N^{1/d}}$ being the average
nearest-neighbor distance.

This immediately gives:  \be P(\lambda)=\frac{d V_d \xi/\langle r
\rangle}{\lambda} [-\xi \rm {log}(\lambda)/\langle r
\rangle]^{d-1}e^{-V_d [-\xi \rm {log}(\lambda)/\langle r \rangle]^d}
. \label{lambda_dist}\ee

Thus, we see that this mechanism yields a $1/\lambda$ distribution,
up to logarithmic corrections, which may be of importance if the
relaxation is probed over enough decades in time. Related analysis
leading to approximately logarithmic relaxations was made in Refs.
[\onlinecite{vaknin,zvi:4,zvi_aging3}], where the relaxation rate was assumed to be the exponential
of a smoothly distributed variable.

Ref. [\onlinecite {Exp_Mat}] analyzes this problem more carefully,
and finds the exact distribution of relaxation rates, for any
spatial dimensions, in the low density limit.
It is shown that the resulting distribution is approximately $P(\lambda) \sim 1/\lambda$,
but in dimensions higher than one there are logarithmic corrections similar to those of Eq. (\ref{lambda_dist}), but with different numerical coefficients. The real-space renormalization group method used also allows to find the structure of the eigenmodes, which show interesting localization properties: all eigenmodes are localized, but with size diverging as one goes to vanishing eigenvalues, as $\sim e^{C |\log^d (-\lambda/2)|}$, with $C$ a constant. In one dimension, for example, this would yield a power-law relation.

An interesting and relevant question addresses the issue of how this
picture is changed when one considers tunneling not only of a single
particle, but simultaneous quantum tunneling of a number of
electrons \cite{manybody1,manybody2,manybody3,manybody4,manybody5,manybody6,manybody7,efros_SCE}. Obviously, the associated timescales would be longer,
since now one has to replace $e^{-r/\xi}$ by $e^{-\sum_j r_j/\xi}$,
where $r_j$ are the tunneling distances of each of the particles
involved in the many-particle tunneling process. In principle there are many possibilities to connect the initial and
final states, but the one where the sum over distances is minimal would be the dominant one.
There should be quantitative changes in the form of $P(\lambda)$, namely, in the
corrections to the numerator of Eq. (\ref{lambda_dist}). Since
estimating the numbers for single-particle processes (see section
(\ref {timescales})) yields too short relaxation times, we are led
to believe that these processes must be considered. A step
in that direction was taken in Ref. [\onlinecite{galperin}].
However, the temperature dependence which they estimate is not
consistent with the rather temperature independent results discussed
above. In our view, the question of many-particle tunneling still deserves further attention.

\section {Aging experiments}

The first experiments showing slow relaxations in the electron glass
were performed by Monroe et \emph{al.}  \cite {monroe1} in 1987, who looked at the
time-dependence of the capacitance of a sample of GaAs, after
injection of excess charge.

Later, Ovadyahu and collaborators have shown remarkable aging
behavior (a concept to be shortly explained) in a series of extensive experiments done mainly on InO
samples, in a field effect transistor setup \cite
{zvi_aging1,vaknin,zvi_aging3,zvi_stress,zvi:TempProtocols,zvi_extrinsic}.
In the following, we review the basic experimental procedure used in
the experiments, and clarify a common confusion between two
different kinds of aging experiments used by the scientific
community. We then go on to review the existing theoretical models
used to explain the results.

 \subsection{Aging protocols}
\label{aging_protocols}

Aging is the general phenomenon related to a relaxation which
depends on the 'age' of the system: in some experiments, the system
is quenched to a low temperature phase at time $t=0$, and the
response $R$ of the system to a perturbation applied at time $t_w$
is tested at time $t$. If $R$ depends
explicitly on both $t$ and $t_w$, and is not only a function
of $t-t_w$, the system is referred to as "aging" over time, since its properties are not translationally invariant
with respect to time.

Indeed, experiments done on spin glasses
\cite{spinglass_aging1,spinglass_aging2,spinglass_aging3,spinglass_aging4}
have shown such "aging" behavior. In this context, the above
protocol is known as a thermoremanent magnetization (TRM)
experiment.

Another protocol, which we will focus on, follows the following
procedure, illustrated in Fig. \ref{protocol}:

\begin{figure}[h]
\begin{center}
$\begin{array}{c@{\hspace{0.00in}}c} \multicolumn{1}{l} {\mbox{}} &
\multicolumn{1}{l}{\mbox{ }} \\ [-0.0cm] \epsfxsize=3.5in
\hspace{0.26 in}\epsffile{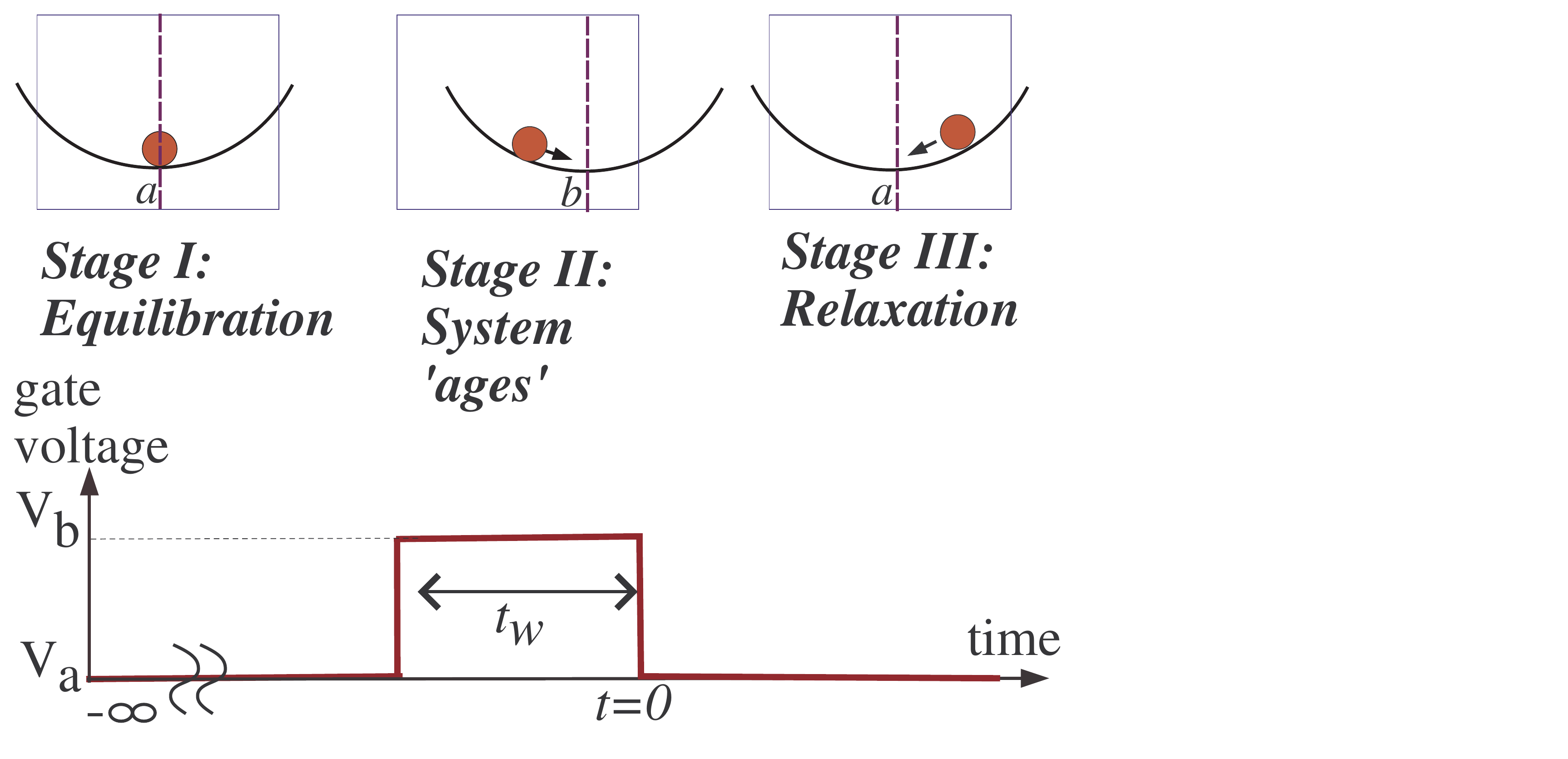} \\
\hspace{-0.7 in}\epsfxsize=3.6in \epsffile{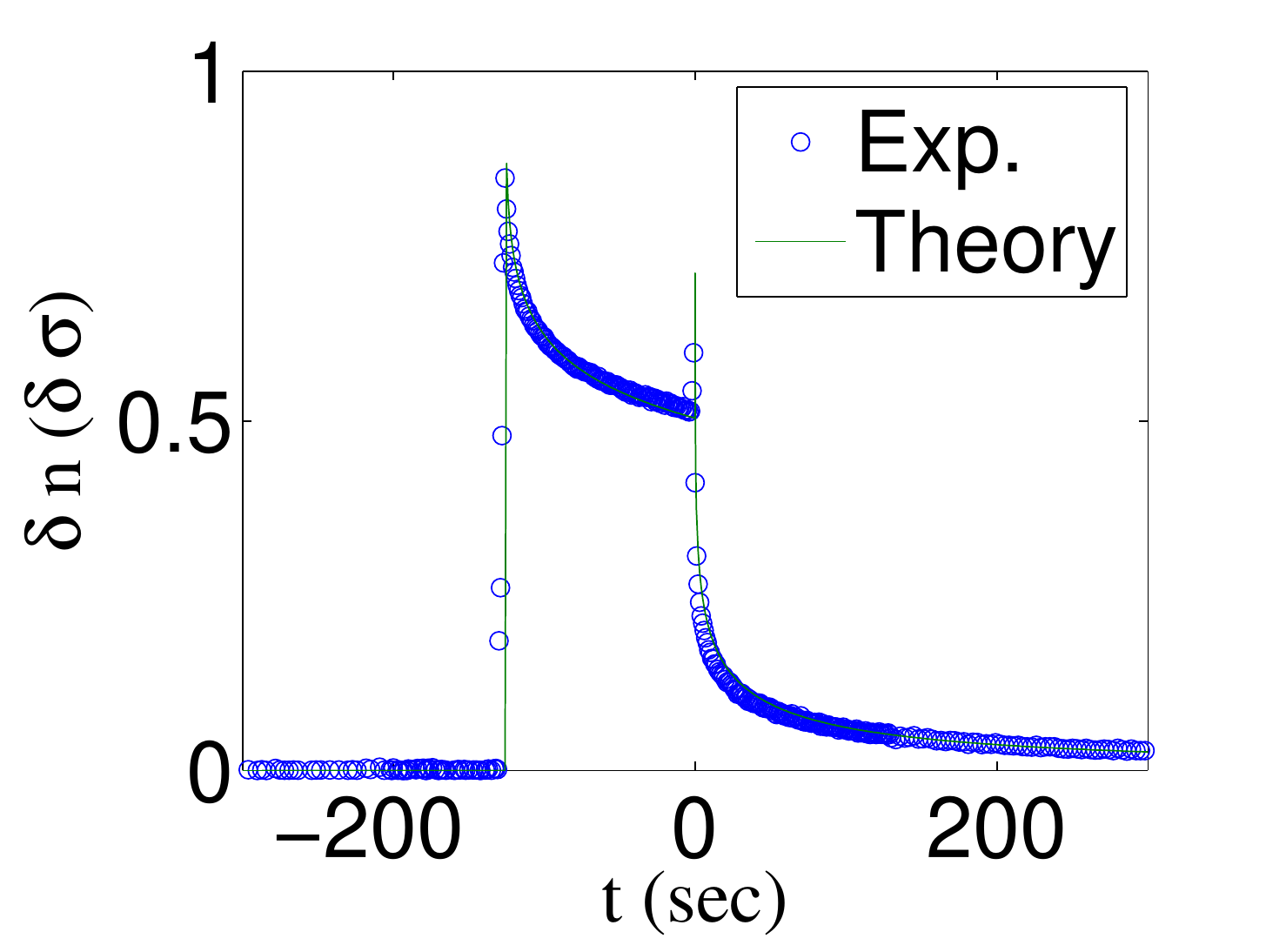} \\
\end{array}$
\end{center}

\vspace{0 cm} \caption{A schematic description of the different
stages of the IRM aging protocol, taken from Ref.
[\onlinecite{amir_aging}].The lower part shows the experimental result (circles),
courtesy of Z. Ovadyahu, and theoretical predictions of the model (solid line), discussed in section (\ref{aging_theory}). Stage II of the experiment shows exactly the same experimental data as Fig. \ref{log_plot}, \emph{i.e.}, at this stage the conductance relaxes logarithmically with time. In stage III, the relaxation turns out to depend only on the ratio $t/t_w$.
\label{protocol}}
\end{figure}

1) The system is let to equilibrate for a long time $t_1$ at the low temperature
phase (not necessarily reaching the true equilibrium).

2) A perturbation is applied for a time $t_w$.

3) A time $t$ after the perturbation has been switched off, the
physical observable $f(t,t_w)$ is measured.

It is important to note that in practice, in most cases the system does not fully equilibrate during the first stage, and in that case this timescale can have implications on the observables in later stages.

The main experimental result for electron glasses are as follows:
During stage II, the relaxation is well described by a logarithm, while in stage III excellent data collapse is obtained
when time is rescaled according to $t_w$, in other words, the relaxation $f(t,t_w)$ in fact only depends on the ratio $t/t_w$.

\begin{figure}[b!]
\includegraphics[width=0.8\textwidth]{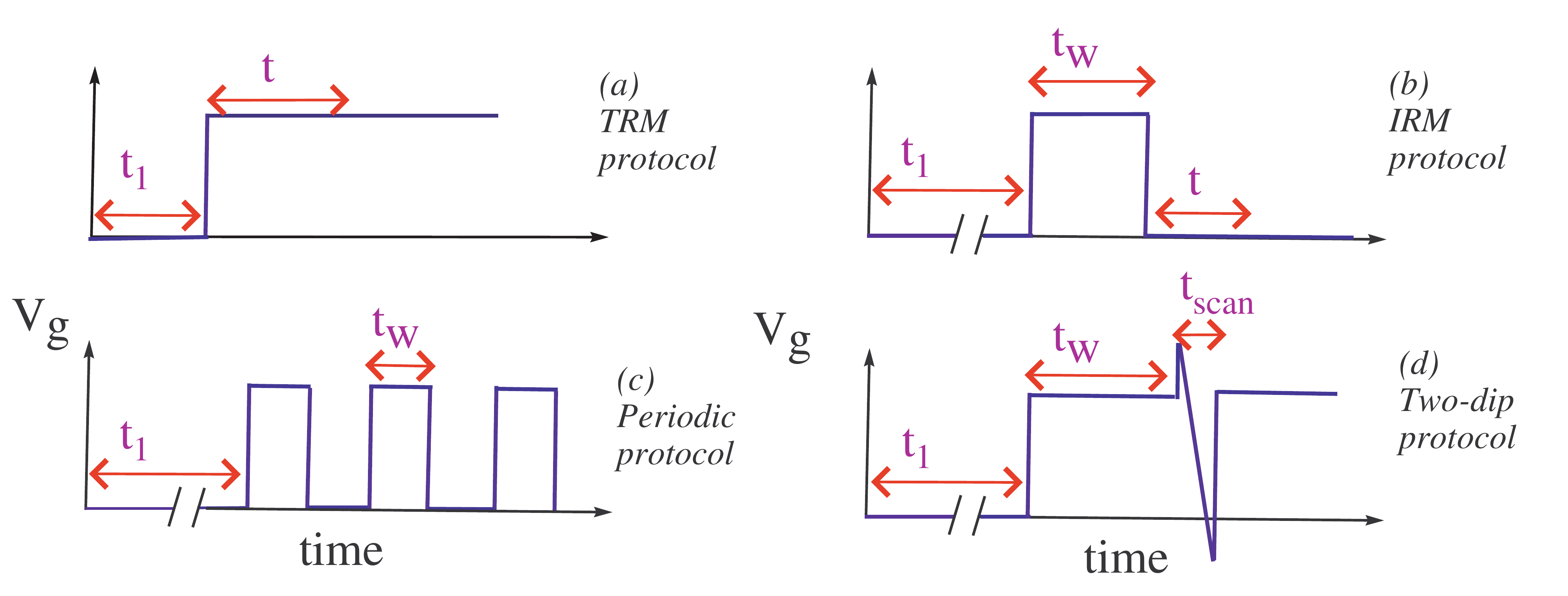}
\caption {Schematic illustration of different protocols in use, relating to the aging and memory effects in electron glasses. In all protocols, at a given time (the crossing point of the $x$ and $y$ axis, in the figure), the temperature is quenched to reach the glass phase. (a) describes the TRM protocol, only recently performed in the context of electron glasses \cite{grenet_2009}, but extensively used in spin glass experiments \cite{spinglass_aging1,spinglass_aging2,spinglass_aging3,spinglass_aging4}. In this protocol, one waits a time $t_1$, and perturbs the system (in the electron glass, typically by changing the gate voltage). A time $t$ later, the response of the system (\emph{i.e.}, the conductance) is measured. If $t_1 \gg t$, one obtains a logarithmic relaxation of the conductance, for times $t$ smaller than the cutoff $1/\lambda_{\rm{min}}$ of the relaxation rate distribution. Such a relaxation is shown in Fig. \ref{log_plot}. The important point, however, is that generally one can also consider $t_1 < t$, which distinguishes this from the IRM protocol. The measured conductance will depend explicitly on both variables $t_1$ and $t$, and in a non-trivial way, which is not well understood. (b) describes the IRM protocol, in which $t_1$ is assumed much larger than other experimental timescales involved (in typical experiments, it is of the order of a day). After the time $t_1$ one changes the gate voltage, for a duration $t_w$ (the 'waiting-time'). A time $t$ after the gate voltage is returned to its initial value, the conductance is measured. Here, the ratio $t/t_w$ can be much smaller or much larger than unity, as long as both are still small compared to $t_1$. Fig. \ref{protocol} shows the experimental results of a typical experiment, and a curve showing the theoretical prediction. The theoretical analysis for this protocol is performed in section (\ref{IRM_theory}). (c) describes a generalization of this, where a periodic square pulse gate voltage is applied, and the conductance is measured throughout the experiment. Unlike most systems, here the response to a periodic signal is not periodic, since the whole experiment is essentially still in the transient period, before the periodic regime is maintained. Fig. \ref{generalized_aging} shows the result of such an experiment. The relevant theoretical analysis is performed in section (\ref{generalized_theory}).  It is assumed that $t \ll t_1$. (d) describes the two-dip protocol: here, the gate voltage is changed at some time, and the system begins to relax logarithmically to its equilibrium. A time $t$ later, a scan of conductance versus gate voltage is made. It turns out that as the system equilibrates at some gate voltage $V$, it 'digs' a dip in conductance at this gate voltage. In this protocol a new dip begins to form at the new gate voltage, as the old one is erased over time, which gives the protocol its name. The protocol is used to demonstrate the memory effects, and also to quantify the timescales involved. It is described in section (\ref{two-dip}), and analyzed qualitatively from the theoretical point of view in section (\ref{anomalous_theory}). Also here it is assumed that $t \ll t_1$. } \label
{exp_protocols}
\end{figure}

In the spin glass context, this type of experiment is known as an
isothermal remanent magnetization (IRM) experiment. Fig. \ref{exp_protocols} illustrates schematically the two protocols described above (TRM and IRM), as well as two other protocols which will be discussed later on.

In a sense, the
IRM experiment is much simpler, since the system is presumably
close to equilibrium at all times during the experiment, provided
the perturbation is small such that we are in the linear-response
regime. Recently, it has been suggested that calling this type of
experiment an aging experiment is misleading, exactly for this
reason \cite {grenet_2009} (since this experiment does not show that the system properties depend on the time from the thermal quench). As we shall later see, understanding the
slow relaxations also for this experiment is actually quite
involved, and presents some unsolved questions.

We shall now review the experimental results obtained for the
different systems.

 \subsubsection{Indium Oxide}

Pioneering experiments using the above $IRM$ protocol, in the context
of electron glasses, have been performed mainly on Indium Oxide, by
Ovadyhau et \emph{al.} In Ref. [\onlinecite{zvi_aging1}] they show that "full"
(also called simple) aging is obtained: the function $f(t,t_w)$
depends only on the ratio $t/t_w$, \emph{i.e.}, $f(t,t_w)=g(t/t_w)$. For short times, they observe
that the function $g(t/t_w)$ depends on the logarithm of its
argument. These properties will be elucidated in the next
subsection. It turns out that this is by no means an accident
related to the peculiar properties of InO, and occurs also for
various other systems. Qualitatively similar results are obtained
for crystalline samples and amorphous samples, demonstrating the
broad applicability of the results. Ref. [\onlinecite{zvi_aging3}] discusses
 the deviations of $f(x$) from a logarithmic dependence,
and also observe deviations from full aging for large enough gate
voltages. In Ref. [\onlinecite{zvi_stress}], Orlyanchik et \emph{al.} use the same protocol
but perturb the system by an electric field (which they termed 'stress aging'), and not a gate voltage. This did
not modify the full aging behavior, for not too large fields.

\begin{figure}[b!]
\includegraphics[width=0.8\textwidth]{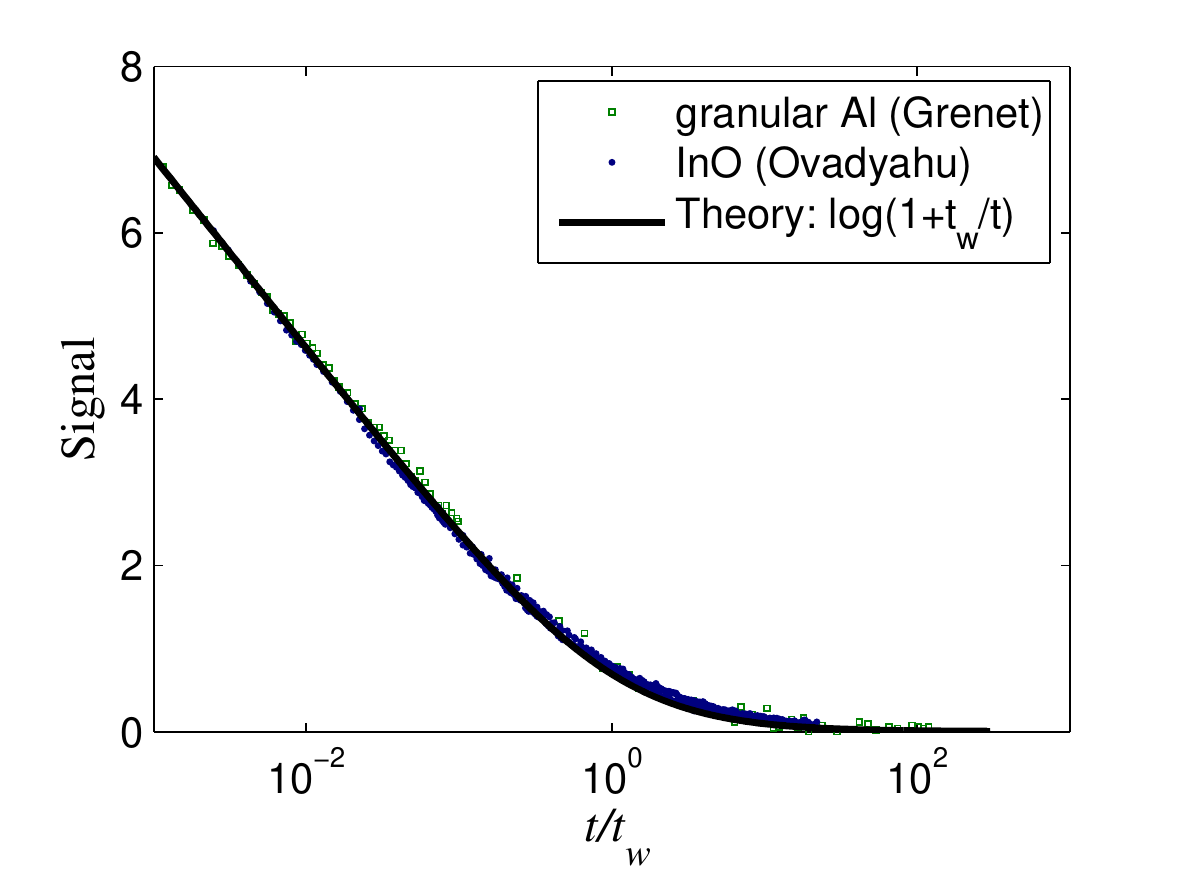}
\caption {Comparison between the theoretical prediction of Eq.
(\ref{aging_eq}) and various experiments. No fitting parameters are
used. Data courtesy of Z. Ovadyahu and T. Grenet.} \label {aging}
\end{figure}

Fig. \ref{aging} demonstrates this scaling behavior for a number of
cases discussed in the following, by showing data collapse when the
time axis is rescaled by $t_w$. The theoretical curve will be
discussed in section (\ref{aging_theory}).

 \subsubsection{ Granular Aluminum}

 Extensive experimental work using the above $IRM$ protocol has been
 performed on granular Aluminum by Grenet et \emph{al.} \cite{Grenet1, grenet_new}. Most of the results show striking
 similarity to those obtained for InO, see Fig. \ref{aging}.
 Some differences occur in the behavior of the two-dip experiment,
 which will be described in section (\ref{two-dip}).

 \subsubsection{Other materials}

Experiments performed on a 2D electron system in silicon by Jaroszy\ifmmode~\acute{n}\else \'{n}\fi{}ski and
Popovi\ifmmode~\acute{c}\else \'{c}\fi{} have shown
slow relaxations and aging \cite{silicon_glass, popovic2,popovic3}. They experimentally observe
a transition between a phase where aging effects exist (dependence
on $t_w$) to one at higher densities where no such dependence exists. In the aging regime, they find full aging only below another critical density. While very
interesting by its own right, it seems that the underlying mechanism
is different than the one which occurs for InO and Al, which is the
one we focus on in this review, and therefore we shall not elaborate
on this extensive set of experiments. Borini et \emph{al.} \cite{Borini}
did observe full aging in porous silicon, at room temperature, which
seems to fit the theoretical framework that we shall put forward in
section (\ref{aging_theory}), see Fig. \ref{aging}.

In the group of A. Goldman, experiments performed on thin films of
bismuth or lead showed logarithmic relaxations in the conductance
after a change in gate-voltage, which relaxed to a different
conductance value, depending on the value of the gate voltage
\cite{goldman2}. They also observed an anomalous field effect
\cite{goldman1}, which will be discussed in section
(\ref{anomalous}). It would be interesting to see the behavior of
this system when the IRM protocol is applied.

Recent experiments have shown that Nickel samples can also exhibit
slow relaxations, with behavior which seems to be similar to that of
granular Aluminum and InO \cite{aviad}. In this case, however, a
transition between a ferromagnetic state and a super paramagnetic
state seems to have a substantial effect on the relaxation rates. As
a result, the magnetic field has a striking effect on the relaxation times. Future experimental and
theoretical work in this direction looks promising.

 \subsection{TRM in electron glasses}
\label{TRM_sect} In a recent experiment Grenet et \emph{al.} have
extended the above protocol to the case when the system is only
partially equilibrated before the measurement sequence is performed
\cite{grenet_2009}. This adds another timescale to the problem
$t_1$, the time from the temperature quench to the first voltage
step, as shown in Fig. \ref{exp_protocols}(a). The above IRM
protocol is the case where $t_1 \gg t_w$. It turns out that for
smaller $t_1$, the measurements depend on it. Remarkably, a
'superposition principle' akin to that observed in spin glasses
\cite{superposition} is still maintained: the response to two
voltage pulses is the sum of the responses to each of the them
independently. This has not been explained theoretically, for the
spin glass nor the electron glass, and would be an interesting
subject for future research.

\subsection{Generalized IRM (periodic protocol)}
\label{Generalized}

\begin{figure}[b!]
\includegraphics[width=0.6\textwidth]{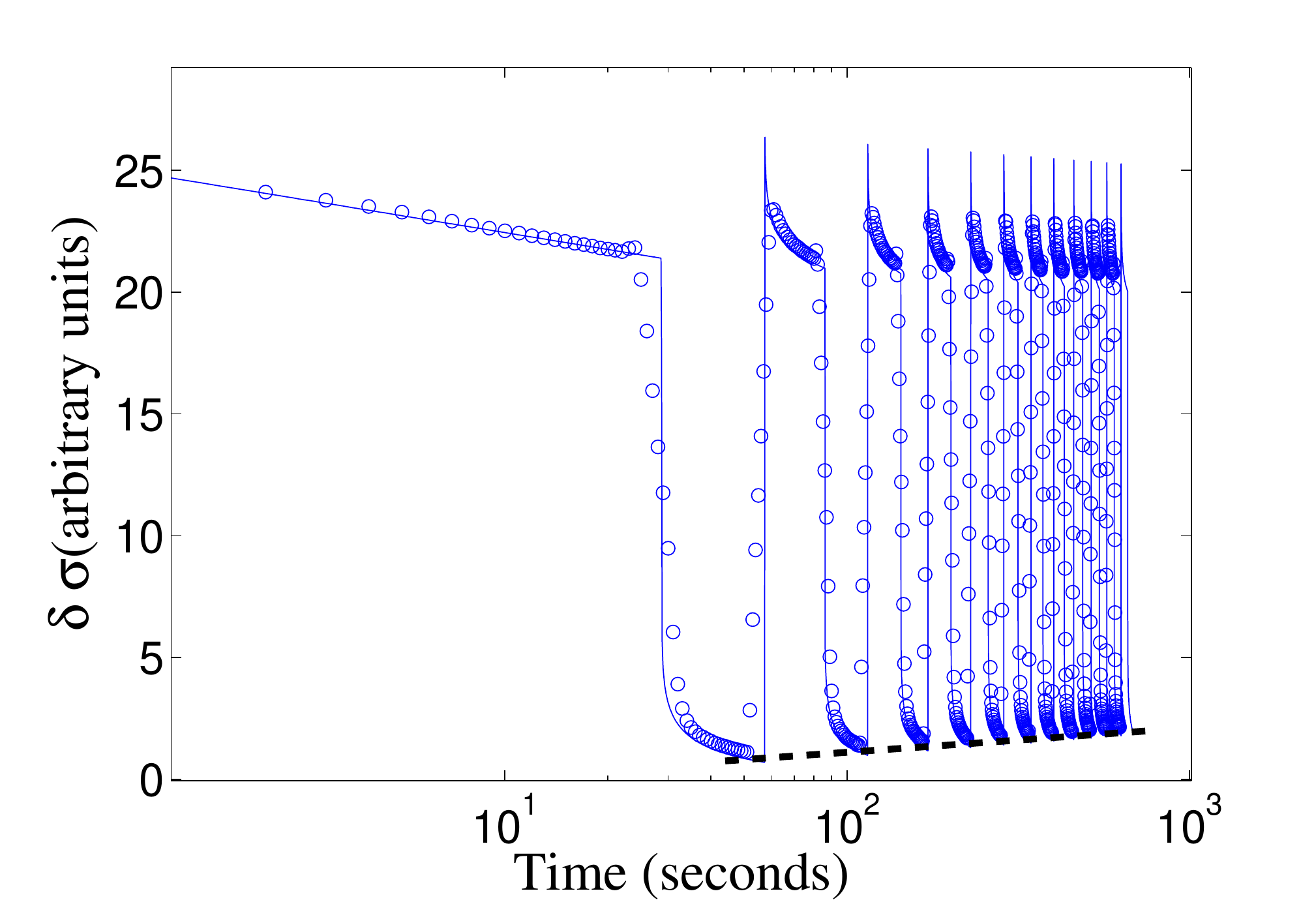}
\caption {Comparison between the theoretical analysis of section
(\ref{generalized_theory}) (solid line) and an experiment (circles) where the sample is
subjected to a periodic perturbation. The dashed line demonstrates the logarithmic increase of the baseline, see Eq. (\ref{log_diverge}). No fitting parameters are
used. Data courtesy of Z. Ovadyahu.} \label {generalized_aging}
\end{figure}

 Another protocol that comes to mind is the generalization of
the IRM protocol to deal with an arbitrary series of pulses. An
elegant experiment is discussed in Ref. [\onlinecite{pollak}], where the
sample has been subjected to a periodic sequence of square pulses, see Fig. \ref{exp_protocols}(c).
Unlike the common scenario, where the response to a periodic signal
is also periodic, after a short transient, here, the whole
experiment is found in the transient period. Fig .
\ref{generalized_aging} shows the result of the experiment, together
with the theoretical prediction, which will be derived and explained
in section (\ref{generalized_theory}).

\section {Why is full aging observed and when does it fail?}
\label{aging_theory}

It is the purpose of this section to present the theoretical models
explaining the aging behavior, and the limitations to obtaining full
aging.

\subsection{Local mean field approach}
\label {meanfield}

In a pioneering work, Thouless, Anderson and Palmer introduced the
TAP equations, to deal with the equilibrium properties of spin glass
\cite{TAP}. A similar approach was used in Ref. [\onlinecite
{coulomb_gap_mean_field}], to deal with the statics of the Coulomb
glass problem. It was shown that by using a local mean-field
approach (which retains the individual identity of each site - not
averaging over the disorder), one obtains the Coulomb gap, a soft
gap in the density-of-states discussed in section
(\ref{coulomb_gap}). The long range nature of the problem gives rise
to a large number of effective neighbors a given electron interacts
with, which gives intuitive justification for the use of the
mean-field approach. This was quantified in various works \cite
{fisher_longrange, kotliar, takashi}, indicating that the relevant
criterion for the validity of the mean-field is that the power
$\alpha$ characterizing decay of the interaction $1/r^{\alpha}$
should be small enough. The results suggest that for Coulomb
interactions in two-dimensions and above one could use the local
mean-field approach.

In Ref. [\onlinecite{amir_glass}], the local mean-field approach was
extended to treat the dynamical aspects.

The equations describing the dynamics are:

\be \frac{d n_i}{dt}= \sum_{j\neq
i}\left(\gamma_{ji}-\gamma_{ij}\right), \label {decay} \ee

with the rates given by Fermi's golden rule\cite{miller_abrahams}:

\be \gamma_{ij} \sim |M_q|^2 \nu f_i (1-f_j)
e^{-\frac{r_{ij}}{\xi}}[1+N(\Delta E)], \label {rates} \ee where
$f_i$ is the Fermi-Dirac distribution, $\Delta E$ is the energy difference before and after the tunneling event,
 $\nu$ is the phonon density-of-state, $M_q$ is the corresponding matrix element and $N(\Delta E)$ is the Bose-Einstein function. For upward transitions
($E_j>E_i$) the square brackets are replaced by $N(\Delta E)$. These
rates may be renormalized due to polaron-type orthogonality effects
\cite{leggett_review}, which were suggested to be of importance in
electron glasses \cite {Ovadyahu_Tdep}.

The Bose-Einstein function expresses the underlying assumption that
the phonons thermalize fast and are found in thermal equilibrium.
$\Delta E$ is the energy difference between the two sites, which is
affected by the Coulomb interactions with all other sites, \emph{i.e.},
$\Delta E= E_j-E_i$, with:

\be E_i= \epsilon_i + \sum_{j \neq i} \frac{e^2 f_j}{r_{ij}} \label
{energies}.\ee

It should be emphasized that in this local mean-field approach the
sites are not assumed to be equivalent, as is clearly seen by the
implicit site indices. This is very different than other
mean-field approaches to the electron glass problem
\cite{pastor,ioffe,pankov2,pankov}.

A characteristic of glasses is the abundance of metastable states,
with energies close to that of the true ground state
\cite{spin_glass}. Also in this case, there are many metastable
states, \emph{i.e.}, electronic configurations for which all the
time derivatives (the LHS of Eq. (\ref{decay})) vanish. This has not
been investigated systematically within this framework, although it
would be interesting to do so. A numerical investigation of the
energy landscape has been performed in a related model through
Monte-Carlo simulations, by Baranovskii et \emph{al.}, where the
metastable states were coined "pseudo ground states"
\cite{baranovskii}. For recent related works see
\onlinecite{garrido, kogan}.

Close enough to a \emph{particular} metastable state characterized
by an electron configuration $\vec{n}_0$, one can linearize the
equations of motion for $\vec{\delta n} \equiv \vec{n}- \vec{n}_0$:

\be \frac{d\vec{\delta n}}{dt}= A \vec{\delta n},
\label{dynamics_meanfield}\ee

where the off-diagonal matrix $A$ is given by the expression
\cite{amir_glass}:

\be { A_{ij}= \gamma^0_{ij}\frac{1}{n^0_j (1-n^0_j)} -\sum_{k \neq
j, i} \frac{e^2 \gamma^0_{ik}}{T}( \frac{1}{r_{ij}}-
\frac{1}{r_{jk}})}, \label{realistic} \ee

with $\gamma^0_{ij}$ the equilibrium rates obeying detailed balance,
see Eq.(\ref{rates}).

 The diagonal matrix elements are given by
$A_{ii}=-\sum_{j \neq i} A_{ij}$, guaranteeing particle number
conservation. The matrix is real but not hermitian, due to the $n_j
(1-n_j)$ term. In section (\ref {noisesect}) we shall give a very
natural explanation for the explicit form of $A$, related to the
properties of the noise via the Onsager relations \cite{amir_noise}:

\be A = \gamma \beta, \label {onsager} \ee where $\gamma$ is a
symmetric matrix describing the transition rates at equilibrium, see Eq. (\ref{rates}), and with the diagonal defined such that the sum of columns vanishes, and $\beta$ is a symmetric matrix whose inverse gives the equilibrium
(equal-time) correlation function between two given sites:

\be \beta_{ij} = \frac{\delta_{ij}}{n^0_i (1-n^0_i)} + \frac{1}{T}
e^2/r_{ij} . \label{beta}\ee

While $A$ is not symmetric, its eigenvalues are nevertheless real: this can be seen using Eq. (\ref{onsager}) \cite{amir_noise}. Near a stable minimum, the eigenvalues must be be negative \cite{amir_glass}, and their distribution will determine the relaxation properties.

\subsubsection{Properties of the relaxation matrix $A$}

In section (\ref{tunnel}), the statistics of nearest-neighbor
distances was analyzed, and was shown to be distributed as
$P(\lambda)\sim 1/\lambda$ up to logarithmic corrections, due to the
exponential nature of the tunneling processes. Looking at the matrix
$A$ of the previous section, we find that the off-diagonal matrix
elements also decay exponentially in the distance. A numerical study
made in Ref. [\onlinecite{amir_glass}] showed that in spite of the more
involved form of the matrix elements (related to the energy
dependence), the above property is still retained. Considering the
toy-model where this energy dependence of the matrix elements is
neglected (which is physically correct when the localization length
is small enough, but is not always the case in experiments), one
obtains a random-matrix class which is defined as follows:

1. $A_{ij}=e^{-r_{ij}/\xi}$.

2. $A_{ii}=-\sum_{j \neq i} A_{ij}$.

As mentioned earlier, the second property arising directly from
particle conservation number.

A heuristic approach to understand qualitatively the emerging
distribution of relaxation rates was made in
Ref. [\onlinecite{amir_glass}]. Later, an \emph{exact} solution was
found, in the low-density case \cite {Exp_Mat}, which confirms the
numerical result described above: up to logarithmic corrections
which depend on the dimensionality and may be of importance if the
experimental resolution is fine enough, the leading order behavior
is a uniform distribution of the \emph{logarithm} of the relaxation
rate, \emph{i.e}., $P(\lambda)\sim 1/\lambda$. This will play a key role in the understanding of the IRM
protocol, to be described next.

\subsubsection{IRM protocol - theory}
\label{IRM_theory}

 In Ref. [\onlinecite{amir_aging}], the $P(\lambda)\sim
1/\lambda$ distribution was the starting point of an analysis of the
IRM protocol. An underlying assumption is that the state of the
system can be described by a vector $\delta {\vec{n}}$, describing
the deviation from an equilibrium or metastable state, and that the
evolution in time is given by:

\be \frac{d \delta {\vec{n}}}{dt}=A \delta {\vec{n}}.
\label{dynamics}\ee Notice that although the above equation is
identical to Eq. (\ref{dynamics_meanfield}), which was derived from
the mean-field approach, here we take a more general approach, where
this equation is the starting point. The distribution mentioned
above is that of the eigenvalues of the matrix.

The crux of the matter is that when a gate-voltage, for example, is
changed, the equilibrium point changes, say from $\vec{n}_a$ to
$\vec{n}_b$, such that the system is instantaneously thrown out of
equilibrium. Relaxation to the new equilibrium is excitation with
respect to the initial equilibrium. Fig. \ref{protocol} describes
the different stages of the experiment schematically.

It is clear that in the stage II, the system is relaxing back to its
new equilibrium, so that in terms of the eigenmodes its deviation
from equilibrium can be written as:

\be \vec{n}(t)=\vec{n}_b+ \sum_q c_q {\vec{b}_q} e^{-\lambda^b_q
(t+t_w)}, -t_w<t<0, \ee where $\lambda^b_q$ are the relaxation rates
of the relaxation matrix $A$ at point $B$, ${\vec{b}_q}$ are the
eigenvectors, and $c_q$ are the weights of the decomposition into
eigenmodes.

Let us assume that the modes contribute, on average, positively and
uniformly to the conductance $\sigma(t)$, \emph{i.e.}:

\be \delta \sigma \sim \sum_{\lambda} |A_\lambda| ,\ee where
$|A_\lambda|$ is the amplitude of the mode $\lambda$ in the
decomposition of $\vec{n}(t)-\vec{n}_b$ into eigenmodes. This is a
central point which will be discussed in section
(\ref{conductance}). We obtain that:

\be \delta \sigma(t) \sim \sum_q e^{-\lambda^b_q (t+t_w)}, -t_w<t<0.
\label {log_relax}\ee

For the large time window $1/\lambda_{\rm{max}} \ll t \ll
1/\lambda_{\rm{min}}$, this gives a logarithmic relaxation:

\be \delta \sigma (\tilde {t}) \sim -\gamma_E-\log [\tilde {t}
\lambda_{\rm{min}}] ,\label{logarithm}\ee with $\gamma_E$ is the Euler
constant, and $\tilde {t}=t_w+t$ is the time from the \emph{first}
change in gate voltage, obviously independent of $t_w$.

Similarly, in stage III, the relaxation is described by:

\be \vec{n}(t)=\vec{n}_a+\sum_q c_q {\vec{a}_q}(e^{-\lambda^b_q
t_w}-1)e^{-\lambda^a_q t}. \label{aging_derivation}\ee

An equivalent sum was given heuristically by Grenet et \emph{al.}
\cite{grenet_new}.

For the IRM protocol, we finally obtain after summing up the series
in the same manner as that done to obtain Eq. (\ref{logarithm})
\cite{amir_aging}:

\be \delta \sigma(t,t_w) \propto \log(t+t_w)-\log(t)=\log(1+t_w/t)
\label {aging_eq}. \ee Fig. \ref{aging} shows the correspondence of
this to the experimental results in 3 different electron glasses.

 We are now
in a position to understand why full aging occurs within this model:
in general, Eq. (\ref{aging_derivation}) predicts a functional
dependence on the parameters $t$ and $t_w$ of the form
$g(t+t_w)-g(t)$. For the specific underlying distribution of
relaxation rate, it was shown that $g(x)$ is logarithmic, which
leads to the full aging of Eq. (\ref{aging_eq}). It can be shown
that the converse is also true, and that $P(\lambda)\sim 1/\lambda$
(associated with a logarithmic $g(x)$) is the \emph{only}
distribution which will give rise to full aging.

It is important to note that in the above analysis it was assumed that the system was initially in equilibrium. As mentioned, in practice it is possible that the system did not fully equilibrate in the first stage of the experiment, and in that case the reciprocal equilibration time $1/t_1$ (see Fig. \ref{exp_protocols}) will replace $1/\lambda_{\rm{min}}$.

\subsubsection{Generalized IRM (periodic protocol) - theory}
\label {generalized_theory}

 It is possible to extend the IRM protocol, to
the case where a more complex sequence of pulses is used. A natural
choice is a periodic sequence of pulses, \emph{i.e.}, the system state is
switched from one value of gate voltage to another, after a $t_w$
length of time. Such a protocol was carried out in detail in
Ref. [\onlinecite{pollak}], as discussed in section (\ref{Generalized}).
We shall now give a theoretical analysis of the experiment, in the
same fashion as was done in the previous section:

At every instance in time, the system relaxes to its current
equilibrium, which we shall denote by $A$ or $B$, depending on the
current value of the gate voltage. Every time the gate voltage is
switched from $A$ to $B$, the vector $\delta{\vec{n}}$ gains an
extra $\Delta \vec{n}$. Similarly, when the voltage is switched from
$B$ to $A$, the gain is $-\Delta \vec{n}$.

Thus, for example, a time $t$ after \emph{two} cycles of gate
voltage changes, the state of the system is described by:

\be \vec{n}(t)=\vec{n}_a+\sum_q c_q {\vec{a}_q}[1-e^{\lambda_b
t_w}(e^{-\lambda^b_q t_w}-1)e^{-\lambda^a_q t_w}]e^{-\lambda_a t}.
\label{generalized_aging_derivation}\ee

Assuming, as before, that points $A$ and $B$ are close enough such
that we can neglect the difference in the eigenmodes and
eigenvalues, the sum can be readily evaluated to give:

\be \delta \sigma \sim \log[1+t_w/(t+2t_w)]+\log[1+t_w/t]. \ee

Writing out the sum explicitly for a general sequence of pulses, we
find the following simple rule: the contribution of a step up is
$\log[\lambda_{\rm{min}}(t-t_{step})]$, and the contribution of step down
is $-\log[\lambda_{\rm{min}}(t-t_{step})]$. Indeed, for the IRM protocol
we obtain $\log(\lambda_{\rm{min}}(t_w+t)-\log(\lambda_{\rm{min}}(t)=
\log(1+t_w/t)$.

Looking at the system state a time $t$ after $N$ pulses, its
conductance is given by:

\be \delta \sigma \sim \sum_{j=0}{^N} (-1)^{j+1}
\log[t+t_w(2j-1)]-\log[t+t_w(2j-2)]. \label{generalize_theory} \ee

To sum up the sequence, it is helpful to sum subsequent pairs of
switches. Looking at the contribution of a pair of switches
occurring a time $t \gg t_w$ earlier, it contributes $\log(1+t_w/t)
\sim t_w/t$. Therefore the response after $N \gg 1 $ pulses is given
approximately by:

\be \sum_{j=1}^N t_w/(2j t_w) = \log (N)/2 = \log[t/(2t_w)]/2. \label{log_diverge} \ee

The fact that this series diverges is profound: it means that we
will have approximately periodic signal, but with a baseline that
diverges logarithmically in time. This is exactly what is
demonstrated in Fig. \ref{generalized_aging}, comparing the
evaluation of Eq. (\ref{generalize_theory}), with no fitting
parameters, with experiments .

The striking result of this analysis, confirmed by the experiment,
is that while we are used to systems responding periodically to a
periodic perturbation after a short transient (e.g: RCL circuits,
mechanical systems etc.), here, the system remains in the
'transient' period for the whole duration of the experiment.

\subsection {Noise in electron glasses}

\emph{1/f} noise occurs in many physical \cite {dutta, Weissman}, as
well as biological and economic systems \cite{heartbeat, economic}.
Electron glasses are no exception, and various experimental studies
have measured $1/f$ noise in such systems \cite{voss,massey,
mccammon, kar, zvi_exp1, non_gaussian, popovicnoise}. An even larger amount of work has been
invested in the theoretical aspects of this problem. The pioneering
works of Shklovskii \cite {shklovskii} were later followed by
various other approaches
\cite{{kogan1},{kozub},{kogan2},{galperin},Shklovskii_noise,
shtengel, yu2}. In a recent
work \cite {amir_noise}, the mean-field framework has been shown to
yield a \emph{1/f} noise in the site occupancies, which has not been
measured until now, but being a local object is a much simpler property than the
conductance noise.

In the following, we shall shortly explain the relation between the
slow relaxations extensively discussed earlier, and the \emph{1/f}
noise. We note that the correspondence to the experimental data should
be done with extra care: since $1/f$ is so common, the experimental
observation of such noise by itself is by no means a confirmation
that the electron glass mechanism yielding this noise is the
dominating one, and additional signatures must be considered.

\subsubsection {Why $1/f$ noise occurs?}
\label{noisesect} Although noise is measured at equilibrium, Onsager
understood in a seminal work that it is generically related to the
way a system returns to equilibrium after a slight perturbation
\cite{onsager}.

In the following, we shall show how this Onsager principle relates
the logarithmic relaxations discussed earlier to the noise in the
system.

\emph{Onsager's regression hypothesis} \cite{LL2} states that the
\emph{equation of motion} of the correlation function $
\label{eq:defcor} \phi_{ij}(t)=\left\langle \delta n_i(t) \delta
n_j(t) \right\rangle$ is obtained by simply replacing $\delta
n_i(t)$ in the equation of motion (\ref{dynamics}) by the function
$\phi_{ij}(t)$: \be \label{eq:phi} \frac
{d\phi_{ij}(t)}{dt}=A_{ik}\phi_{kj}(t) .\ee

To find the correlation function one also needs the initial
conditions:

\be \phi_{ij}(0)=\left\langle \delta n_i(0) \delta n_j(0)
\right\rangle \equiv \beta^{-1}_{ij}. \ee

To find the matrix $\beta$, we write the free energy
\cite{coulomb_gap_mean_field}:

\be F = \sum_i \epsilon_i \tilde{n}_i +\sum_{i \neq j} \frac{e^2
\tilde{n}_i \tilde{n}_j} {r_{ij}}+T\sum_i\left[(\frac{1}{2}+
\tilde{n}_i) \log(\frac{1}{2}+
\tilde{n}_i)+(\frac{1}{2}-\tilde{n}_i) \log
(\frac{1}{2}-\tilde{n}_i)\right],\ee  with $\tilde{n}_i \equiv n_i
-\frac{1}{2}$. The local mean-field equations can be obtained from
the minimalization condition $\frac{\partial
F}{\partial\tilde{n}_i}=0$. Notice that each site contains a
positive background charge of
$\frac{1}{2}$, to keep charge neutrality.

Expanding $F$ near a metastable state (a local minimum), we have:

\be F= F_0 + \frac{kT}{2}\sum_{i,j}\beta_{ij}\delta n_i \delta n_j
,\label {Fexpand}\ee

with $\beta$ given by Eq. (\ref{beta}).

The correlation matrix is proportional to $\beta^{-1}$, since we
have a quadratic free energy:

\be \langle \delta n_i \delta n_j \rangle  =
\left(\beta^{-1}\right)_{ij}. \ee

Regardless of the question of determining the noise spectrum, which
we shall shortly derive, at this stage we can use another theorem
due to Onsager to put the form of the non-hermitian relaxation
matrix $A$ of Eq. (\ref{realistic}) in a more natural context: the
Onsager symmetry principle states that $A \beta^{-1}$ must be a
symmetric matrix. Indeed, in our case $A= \gamma \beta$, with
$\gamma_{ij}$ the equilibrium current between sites $i$ and $j$, as
was mentioned in section (\ref{meanfield}).

Let us proceed to the calculation of the noise spectrum. After
finding $\beta$, we can solve Eq. (\ref{eq:phi}), and obtain the
noise spectrum of the average site occupancy \cite{amir_noise}. This
leads to the expression:

\begin{equation}
\label{eq:sumform}
 \phi_{ii}(\omega) = \sum_{\alpha}
\frac{1}{\beta_{ii}} \left|\psi^\alpha_i\right|^2 \frac{2
\lambda_\alpha}{\omega^2+\lambda_\alpha^2},
\end{equation}

where $\psi^\alpha_i$ describes the amplitude of the $\alpha'th$
eigenmode of the relaxation matrix $A$. The factor
$\frac{1}{\beta_{ii}}$ describes the physically clear fact that only
sites with energies close to the Fermi-energy will contribute to the
noise, since they are "soft" and their occupancy fluctuates
significantly. We therefore find that:

\be \langle \langle \delta n^2\rangle \rangle_\omega  \sim
\frac{1}{N}\sum_{\alpha,i}
  \frac{\frac{2}{\lambda_\alpha}}{1+(\frac{\omega}{\lambda_\alpha})^2} , \label {t_dep}\ee
  where $\langle \langle;\rangle \rangle$ denotes averaging over sites as well as time.

Plugging in the $P(\lambda) \sim 1/\lambda$ distribution of rates,
we finally obtain the noise spectrum:

\be \langle \langle \delta n^2\rangle \rangle_\omega  \sim
\frac{1}{N}\int_{\lambda_{\rm{min}}}^{\lambda_{\rm{max}}} d\lambda
  \frac{\frac{1}{\lambda ^2}}{1+(\frac{\omega}{\lambda})^2} =\frac{1}{N
\omega}\int_{\frac{\lambda_{\rm{min}}}{\omega}}^{\frac{\lambda_{\rm{max}}}{\omega}}
dm
  \frac{1}{1+m^2} . \label {w_dep}\ee

This shows that for $\lambda_{\rm{min}} \ll \omega \ll \lambda_{\rm{max}}$, a
\emph{1/f} spectrum indeed follows for the noise in the average
occupation number.

\section {Memory in electron glasses}
\subsection {Anomalous field effect}
\label{anomalous}

\begin{figure}[b!]
\includegraphics[width=0.4\textwidth]{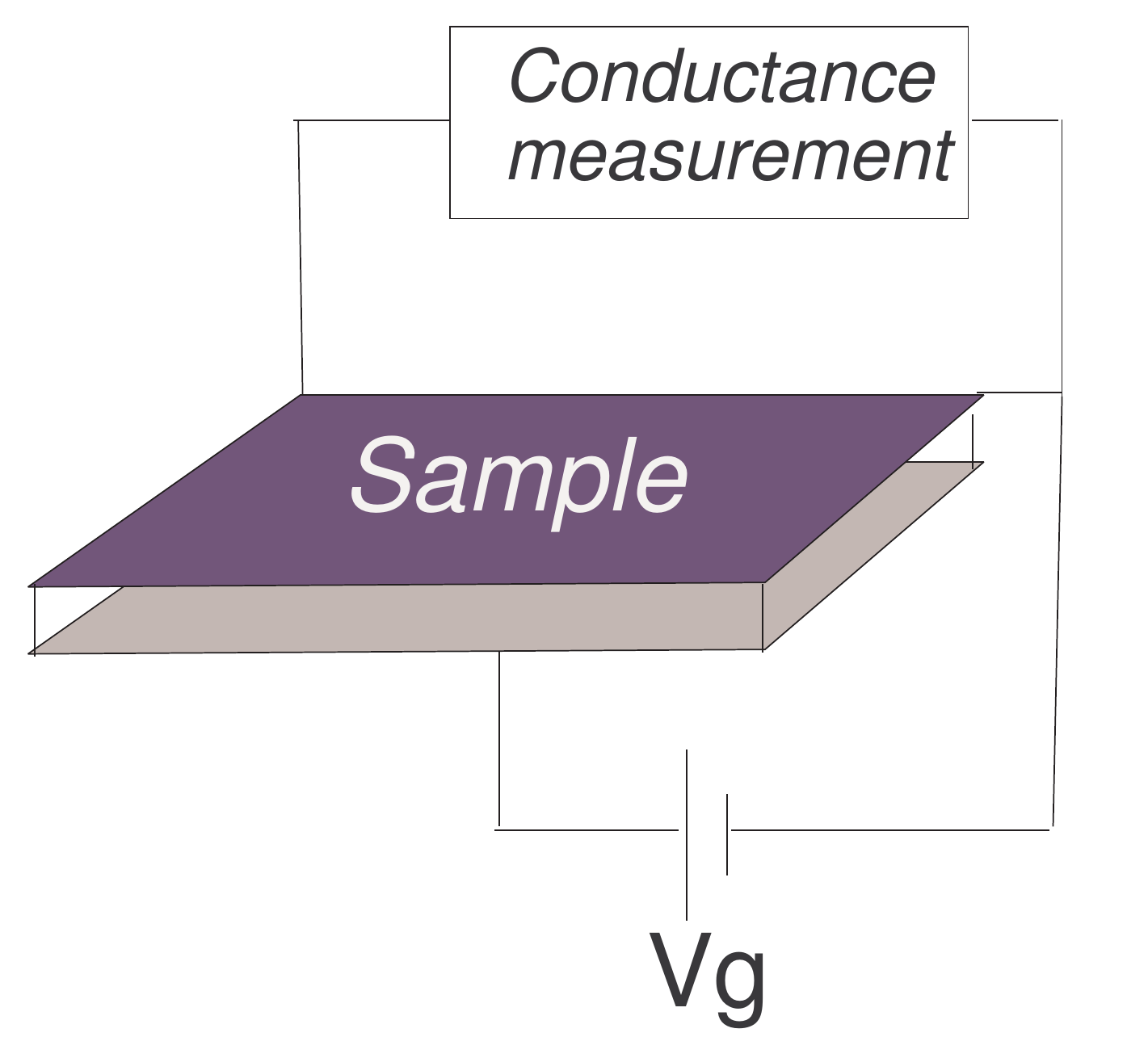}
\caption {Schematic illustration of the field effect transistor
setup. A gate voltage is attached to the sample via a separator, and
the conductance of the sample is measured while changes in the gate
voltage charge the sample, positively or negatively. } \label {FET}
\end{figure}

There is a clear experimental demonstration that for electron glasses perturbing the
system causes the conductance to increase. In 'ordinary'
field effect transistors, where interactions are not as strong,
adding charge carriers to the system by changing a gate voltage (see
Fig. \ref{FET}), will increase the conductance, and taking charge
away will make it decrease. In electron glasses, however, after
full or partial equilibration, changing the charge carriers number in any direction will
make the conductance \emph{increase}. This surprising property is
called the anomalous field effect \cite{zvi_anomalous_misc}. It is
clear that the only thing which distinguishes the particular value
of gate voltage used is the fact that the system was let to
equilibrate in it. Moreover, it is clear that performing the
experiments 'quasi-statically', waiting a long enough time at each
voltage point, will retrieve the normal field effect. Thus, the
voltage scan rate must enter. Fig. \ref{anomalous} demonstrates the
results of scanning at different rates. Intuitively, relaxing to
lower in energy (deeper) metastable states should indeed result in a lower
conductance. The theoretical framework for understanding this
property will be given in section (\ref{anomalous_theory}).

\begin{figure}[b!]
\includegraphics[width=0.6\textwidth]{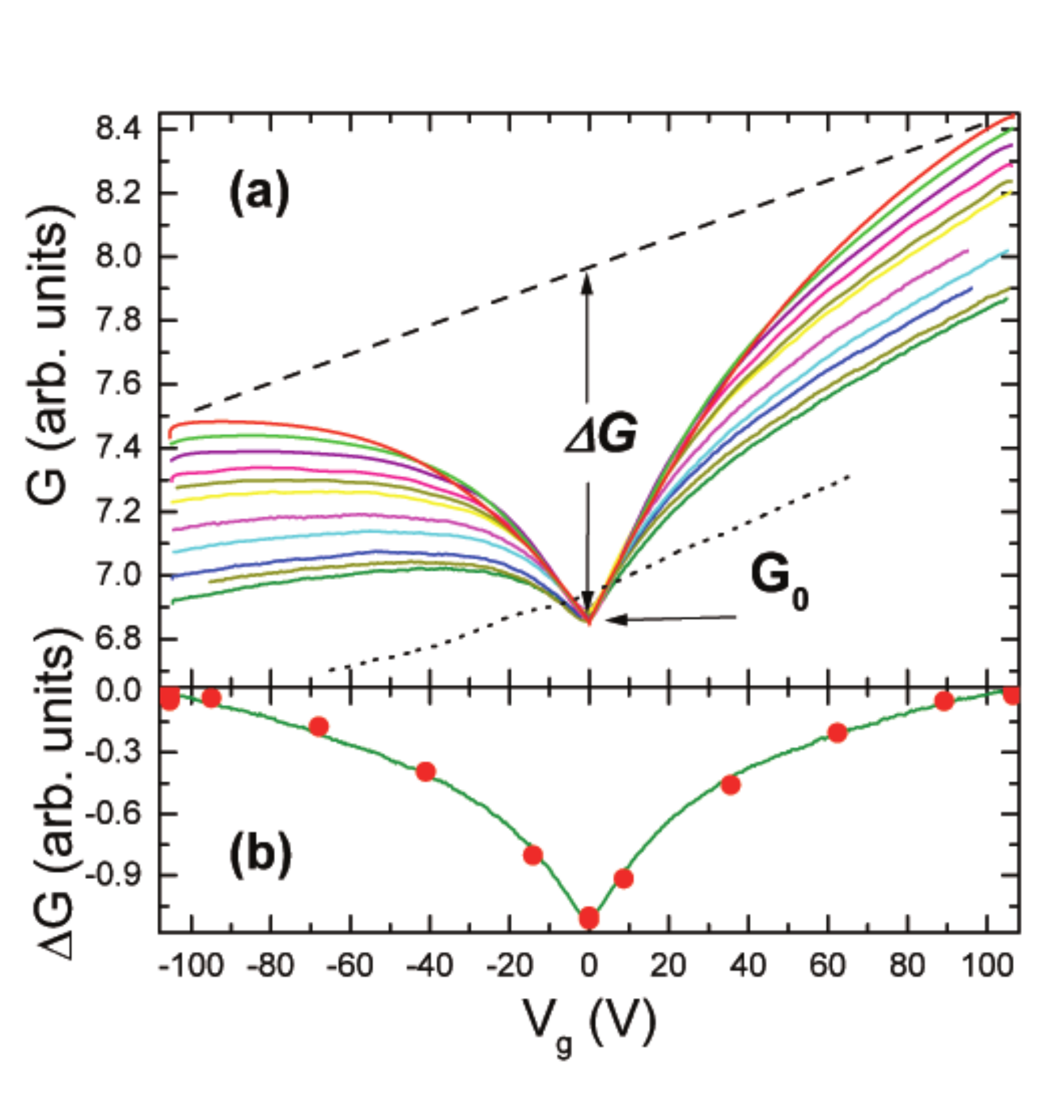}
\caption {The dependence of the conductance on the gate voltage was
checked for various scan rates, in a typical field effect transistor
setup (see Fig. \ref{FET}). In addition to a linear trend, related
to the "normal" field effect, there is a surprising symmetric
component, whose amplitude depends on the scan rate (a). Subtracting
the linear trend and normalizing the amplitude, the rate dependence
is eliminated (b). Figure taken from Z. Ovadyhau, PRB 78, 195120
(2008) with permission from the author. } \label {anomalous}
\end{figure}

The width of the dips (also referred to as cusps) is an important characteristic of the system:
an elaborate set of experiments \cite{extrinsic_misc} have shown
that the cusp shape does not depend on the scan rate (for scan rates
varying over 3 decades), and is independent on disorder and magnetic
field. As such, it manifests an intrinsic property of the system,
which has also been shown to be correlated with the interaction
strength (by showing a strong density dependence)\cite {zvi:2}.

We now go on to describe a slightly more involved experimental
protocol, that can serve as a useful characterization tool to probe
changes in the system timescales.

\subsection {Two-dip experiments}
\label {two-dip}

An important protocol for demonstrating and quantifying the memory
effects in electron glasses in a striking way was developed by
Ovadyahu et \emph{al.} \cite{ziv_disorder, zvi:2}. The experimental
protocol is as follows (see also Fig. \ref{exp_protocols}):

 1. As in the anomalous field effect, the system is let to
equilibrate for a long time ($t_1$ in Fig. \ref{exp_protocols}),
typically of the order of day or several days. The gate voltage is
fixed at a value $V_1$ during this time.

2. At time $t=0$, the gate voltage is changed to a value $V_2$.

3. A time $t_w$ later, a scan of conductance vs. gate voltage is
made.

\begin{figure}[b!]
\includegraphics[width=0.7\textwidth]{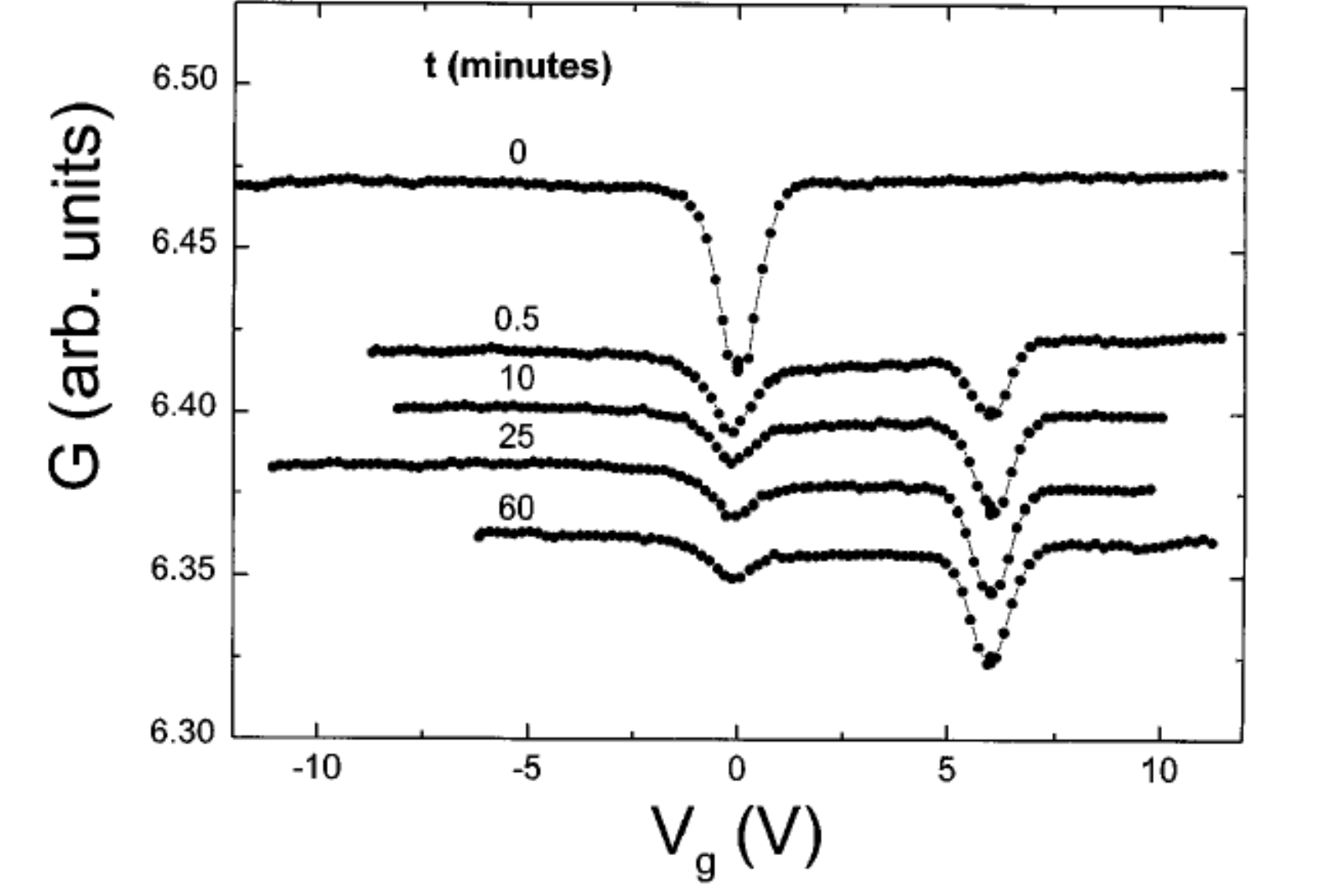}
\caption {A typical experimental measurement for the two-dip
experiment protocol. Initially, the anomalous field effect accounts
for the dip at zero gate voltage, at which the system was
equilibrated. The gate voltage is then shifted to a different value,
and as the system equilibrates a new dip begins to form at the new
value while the initial one is slowly erased. Figure taken from A.
Vaknin, Z. Ovadyahu and M. Pollak, PRL 81, 669 (1998) with
permission from the author. } \label {twodip}
\end{figure}

Clearly, if $t_w=0$, we observe the anomalous field effect mentioned
earlier. However, for finite values of the waiting-time $t_w$, the
system exhibits a striking memory effect: while the dip at the
original gate voltage $V_1$ is gradually erased, a new dip begins to
form at the gate voltage $V_2$. For this reason this important
protocol is termed the 'two-dip experiment'. Fig. \ref{twodip} shows
the result of a typical experiment.

At some time $t_w$, the two dips will be equal in magnitude. This
provides a natural experimental time $\tau$ characterizing the
system. Experimentally, Ref.
[\onlinecite{zvi:TempProtocols}] demonstrates that this timescale
can be consistently measured in various other ways: for example, one may change the value of the gate voltage when the
 system is close to equilibrium, and measure how long it takes for the (logarithmic) change in conductance to decay to half its value at one second. We will discuss the physical significance of these timescales
in section (\ref{anomalous_theory}), and show why the two approaches give similar results.

In Ref. [\onlinecite{Ovadyahu_Tdep}], the temperature dependence of this
timescale was carefully measured. The result appears to be quite
counter-intuitive: as the temperature increases, and one may expect
the dynamics to be faster, $\tau$ actually increases. This was
associated with a quantum effect, reminiscent of the anomalous
temperature effect which occurs in the dynamics of the spin-boson
problem for coupling to an Ohmic bath and for a certain regime of
parameters \cite{leggett2}. However, applying a theory
constructed for a two level system to one as complex as electron
glass is possibly an oversimplification, and this point requires
further theoretical attention. Furthermore, one should make sure that the measured timescale is indeed an intrinsic timescale of the system, not associated with $t_1$, as mentioned in section (\ref{aging_protocols}).

A subtle but important difference between the experiments done on
InO and those on granular aluminum \cite{Grenet1, grenet_new}, relates to the evolution in time
of the plateaus outside the dips (cusps). While in InO these are static
in time, in granular aluminum they show a logarithmic dependence on
time as well. Recently, T. Grenet and J. Delahaye showed this could
be consistently understood in terms of metallic screening in the
thin films \cite{screening}.

\section{ How electron glasses remember - theoretical picture}

\subsection {Connection between the conductance and the occupation number
relaxations}

So far we discussed the relaxations of the occupation numbers. One
still has to explain how this influences the conductance, which is a
much more involved property. Namely, in the analysis we assumed that
the relaxation modes contribute positively to the conductance.
Experimentally, it is indeed clear that the conductance is raised
when the system is perturbed out of equilibrium. However, it is not
obvious theoretically, a-priori, why any perturbation of the
occupancies will tend to raise it. Ref. [\onlinecite{zvi:1}] shows
how this can come about for a non-interacting electronic systems (in addition to giving experimental results). In
the following, we shall give two explanations of this remarkable
property, where the interactions play a central role. Before that,
we shortly review the properties of the equilibrium conductance,
which will be important for the discussion which follows.

\subsubsection{Miller-Abrahams resistor network}
\label{miller_sect} In a seminal paper, Miller and Abrahams showed
that the conductance of a system of non-interacting electrons,
'hopping' between localized states, can be mapped to the
determination of the resistance of a resistor-network, where the
value of the resistor between site $i$ and site $j$ is given by
\cite{miller_abrahams}:

\be R_{ij}= T/[e^2\gamma^0_{ij}], \label {Miller}\ee with
$\gamma^0_{ij}$ the equilibrium hopping rate from site $i$ to site
$j$, see Eq. (\ref{rates}). Detailed balance, assumed to hold at
equilibrium, ensures us that the resistor is well-defined. The
intuition behind this is that well coupled sites will have a good
conductance between them. As one raises the temperature, the effect
on $\gamma_{ij}$ is exponential, and exceeds the linear term in the
numerator. For a mesoscopic sample (or for numerical purposes) it
might be important to deal with the connections to the leads. Ref.
[\onlinecite{amir_vrh}] shows how to incorporate them into the same
framework, with the result that the resistor between a site and the
leads takes the same form as that of Eq. (\ref{Miller}). When all
the energies involved are much larger than the temperature, the
rates $\gamma_{ij}$ can be shown to take the approximate form
\cite{ambegaokar}:

\be \gamma_{ij} \sim \rm{exp} [-\frac{2r_{ij}}{\xi}-
\frac{|E_i-\mu|+|E_j-\mu|+|E_i-E_j|}{2T}] \label{rate_VRH} ,\ee

This is commonly used regardless of the above restriction,
erroneously. In Refs. [\onlinecite {amir_vrh, efros}] it is shown that by
taking interactions on the local mean-field level, this formula remains
correct, but with $E_i$ renormalized due to the Coulomb interactions
between sites.

Determining the resistance of the complete network is still not an
easy task. Extensive work has been done on this problem using
percolation theory methods \cite{efros}, some using the concept of percolation in phase-space \cite{manybody6,manybody7}. A direct, 'brute force'
method is the numerical determination of the resistance, via
Monte-Carlo simulations \cite{amir_vrh}. In the following subsection
we show a heuristic approach due to Mott, later extended by Efros
and Shklovskii to deal with the interacting problem, that finds the
leading order behavior.

\subsubsection{Variable Range Hopping}
\label{vrh_sect} In 1969, Mott considered the problem of hopping
conductance, \emph{i.e.}, the conductance occurring through phonon assisted
jumps (hops) between localized states. In a brilliant analysis, Mott
realized that at low enough temperatures it is beneficial for the
electrons to hop not to the nearest-neighbor, but to neighbors more
further away \cite{mott}. This is in spite of the
exponential-in-distance penalty, in order to reduce the
exponential-in-energy penalty. The temperature determines the scale
for the exponential-in-energy penalty, and thus at extremely low
temperature the electron tunnels over a large distance, to find a
site close to it in energy, see Eq. (\ref{rate_VRH}). The result is
that:

\be \sigma \sim e^{-\left(\frac{T_0}{T}\right)^{-\frac{1}{d+1}}},
\ee where $d$ is the dimension of the system, and $T_0 \approx
\frac{1}{\nu \xi^{d}}$, with $\nu$ the density-of-states and $\xi$
the localization length (taking the dielectric constant to be
unity).

\subsubsection{Coulomb gap}
\label {coulomb_gap}

About 4 decades ago, it was understood that Coulomb interactions can
have profound effects on the density-of-states (DOS) of electronic
systems near the Fermi energy, namely, leading to the vanishing of
the DOS at the Fermi energy \cite{pollak_} and the emergence of a
'soft' gap near it \cite {ES_gap}, suggested by an analysis by Efros
and Shklovskii. Since then, a large number of analytical \cite
{efros, efros_SCE, ioffe, pankov, raikh}, numerical \cite
{{numerics:2},{numerics:1},{numerics:3},{numerics:4},{numerics:5},
{numerics:6}} and to a lesser extent experimental \cite
{{coulomb_gap_experimental}, {butko}} approaches have dealt with
this interesting problem. The current understanding is that there is
a power-law DOS at equilibrium, with exponent depending on the
dimensionality. Since it is a direct result of the Coulomb
interactions, it is called the Coulomb gap.

\begin{figure}[b!]
\includegraphics[width=0.6\textwidth]{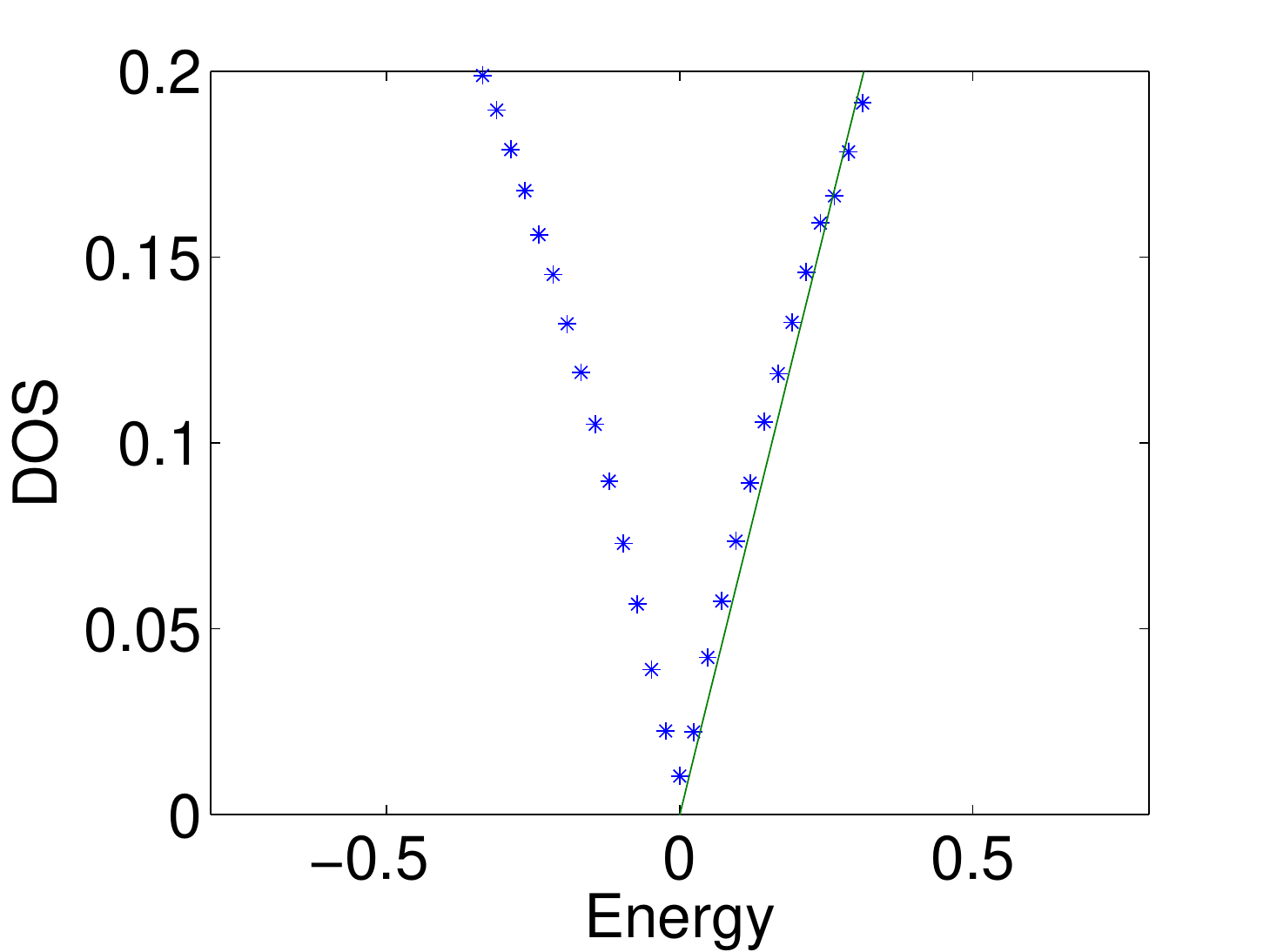}
\caption {Comparison between the local mean-field approach
\cite{amir_glass} and the Efros self-consistent equation approach
\cite{efros_SCE}, yielding a linear Coulomb gap (solid line) in
two-dimensions, with slope $\sim \frac{2}{\pi e^4}$ . The figure is
taken from Ref. [\onlinecite{amir_vrh}]. \label{efros_meanfield}}
\end{figure}

For the current work, we shall not attempt to review the literature
on the subject. The only ingredient that we shall take away from
this for the following, is the existence of a soft-gap near the
Fermi energy. Surprisingly, this can also be understood from the
local mean-field theory discussed in section (\ref{meanfield})
\cite{coulomb_gap_mean_field}. In two-dimensions, Ref.
[\onlinecite{amir_vrh}] shows that the gap obtained by the local
mean-field approach is consistent with that predicted by Efros
\cite{efros_SCE} and Raikh \cite{raikh} by a self-consistent
equation approach (taking only single electron transitions into
account). Fig. \ref{efros_meanfield} compares this 2D result with
the Efros prediction of $\frac{2|E|}{\pi e^4}$. It would be
worthwhile to show how this comes about by analytically solving the
local mean-field equations, which has not been done to this date.

\subsection{Why the conductance increases out-of-equilibrium : Coulomb gap}
\label{conductance}

 It is plausible that when pushed out-of-equilibrium,
the Coulomb gap is washed out. It is interesting and experimentally
relevant, to ask how the gap is recovered in time. Experimentally,
most of the gap would be recovered in very short times, as is
analyzed extensively by Tsigankov et \emph{al.} \cite{pazy}. At long times,
there could be two possibilities: the first, is that all metastable
states manifest the same Coulomb gap, and that the complete recovery
of the Coulomb gap to each of them is a slow process, for energies
close to the Fermi energy. This approach was supported by Yu by
generalizing the Efros self-consistent approach to the
time-dependent domain \cite {yu}. The second possibility is that as
time goes on, the system probes different metastable states, each of
which possseses a slightly different Coulomb gap, which 'deepens' as
the systems relaxes by going into deeper (lower in energy)
metastable states. Work in this direction was done in
Ref. [\onlinecite{pazy}].

The idea is now as follows: it is plausible that the DOS near the
Fermi energy is crucial for the conductance properties. This was
used by Efros and Shklovskii in a seminal work , to show how the
Mott picture for variable-range-hopping, briefly discussed in
section (\ref{vrh_sect}), should be modified \cite{ES_gap}. Although
the use of the single particle tunneling DOS in the conductance
calculation is not justified \cite {Pollak_1_over_2}, their
prediction was confirmed in a huge number of experimental systems
\cite {ES_VRH_exp}, functionally but not necessarily quantitatively.
In a recent work \cite {amir_vrh}, it was shown that by following
the local mean-field approach, one can self-consistently take the
interactions into account, and essentially verify the
Efros-Shklovskii picture. Thus, the Coulomb gap reduces the
conductance, by reducing the density of available states near the
Fermi energy. The picture that emerges is that through the Coulomb
interactions the system 'digs' a hole in the DOS, and makes its own
conductance diminish. Therefore, by kicking the system out of
equilibrium, we enhance the conductance, as we assumed earlier in
the analysis of the IRM protocol, by taking the contribution of the
modes to be positive. The problem that remains is the understanding
of the timescales: as we mentioned, various references suggest that
the Coulomb gap is created on fast timescales of the order of the
Maxwell time \cite{efros} (although, as mentioned, some works
indicate that at energies close to the Fermi energy, \emph{i.e.},
the bottom of the gap, the Coulomb gap takes a long time to form
\cite{yu}). The next section will give a possible solution, that
shows how this effect can be understood on more general ground,
beyond the single particle relaxations.

 \subsection{Why the conductance increases out-of-equilibrium : long time relaxations} \label {polaron}

In the presence of Coulomb interactions, slow relaxing modes can
modify the energies of other sites, and thus influence the
conductance even if the modes are isolated, and the tunneling
between them and the rest of the sites is negligible (since it is
exponentially suppressed). This picture also explains why one may
talk about an out-of-equilibrium conductance: there can be a
separation of timescales between the (relatively quick)
equilibration time of the current carrying sites, the 'backbone',
and the isolated, slowly relaxing modes outside it. When the slow
modes have not equilibrated with the rest of the system, strictly
speaking the system is out-of-equilibrium. Nevertheless, there is a
well defined conductance that can be experimentally measured,
defined by the internally equilibrated sites belonging to the
backbone, which have reached quasi-equilibrium in the presence of
the slowly changing potential created by the slow modes
(\emph{i.e.}, their statistics is described by the Fermi-Dirac
distribution). In this sense, the combination of the long ranged
Coulomb interactions and the localized modes explains why the
conductance returns slowly to its equilibrium value. This is not
enough, however, to explain why it does so \emph{monotonically}, as
is clearly seen experimentally in the IRM experiment, for example.

In order to understand this, a simple argument has been put forward
by Ref. [\onlinecite{galperin}]. In the following we give our
version of the argument.

\begin{figure}[b!]
\includegraphics[width=0.6\textwidth]{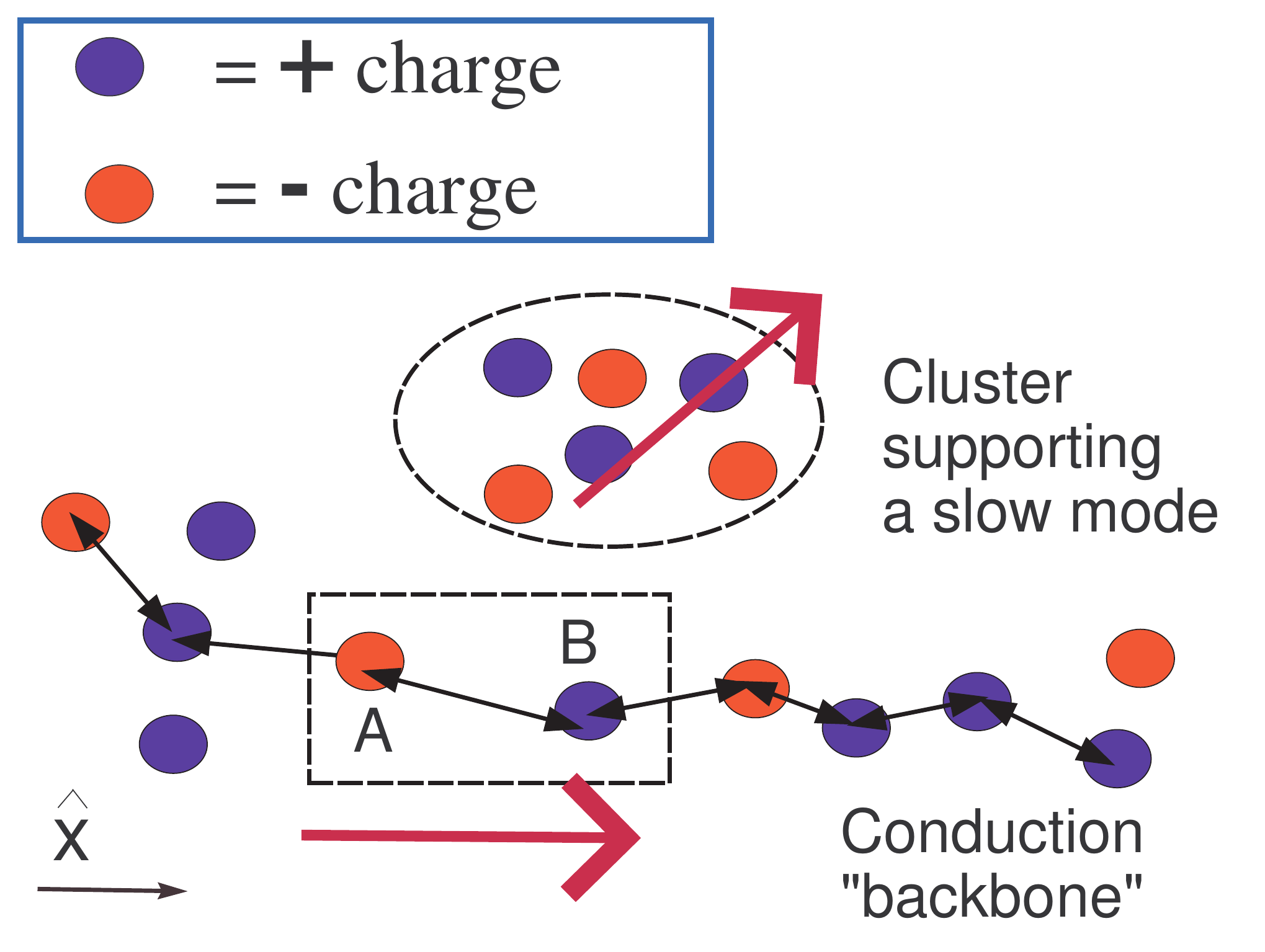}
\caption {Schematic illustration used to demonstrate why a
relaxation of a mode outside the conduction "backbone" contributes
negatively to the conduction, on average. The dotted (purple) arrows denote the dipole moments of the cluster and of two sites in the backbone. These will prefer to point in opposite directions, on the average, and thus while a cluster relaxes it will tend to \emph{raise} the energy difference between the two sites in the backbone, \emph{on average}, thus making the conductance lower. \label{polaron_fig}}
\end{figure}

An essential ingredient is the separation of timescales between
those associated with hopping between sites which contribute
significantly to the conduction (by definition "fast" processes),
and clusters of sites which support "slow" modes: Ref. [\onlinecite
{Exp_Mat}] shows that isolated clusters lead to an abundance of
slow modes with a distribution approximately given by
$P(\lambda)\sim 1/\lambda$, but this is not the only mechanism
leading to slow modes: Coulomb interactions will give rise to modes
related to the tunneling of many electrons, related to the work of
Kozub et \emph{al.} \cite{galperin}. The contribution of these clusters to
the conductance is not negligible, however, since they affect the
energies of the sites associated with the fast processes (from now
on we refer to these as the "backbone"), through the Coulomb interactions. One might expect that the contribution will be
effectively random: sometimes they will give rise to an increase and
sometimes a decrease. It turns out that \emph{on average}, however,
they give rise to a \emph{positive} contribution. We now go on to
the crux of the matter: let us consider a relaxation mode associated
with an isolated cluster some distance away from the backbone, as shown in
Fig. \ref{polaron_fig}.

Since we have a separation of timescales, it is still justified to
think of the Miller-Abrahams resistance network associated with the
conduction backbone (see section (\ref{miller_sect}), which is
presumably in equilibrium, but with the energies of the sites in it
affected by the current configurations of the slowly relaxing modes
outside the backbone, which have not yet fully equilibrated. Let us
consider the value of the resistance associated with two close
sites on the backbone, $A$ and $B$ in Fig. \ref{polaron_fig}. The
resistance between them is given by Eq. (\ref{rate_VRH}), showing
that the dependence on energy is:

\be R \sim \rm{exp} [\frac{|E_A-\mu|+|E_B-\mu|+|E_A-E_B|}{2T}]
\label {R} \ee

The claim is that modes that relax in energy will tend to make the
value of the resistor larger (on average) as they relax. Let us
assume that $E_A<\mu$ and $E_B>\mu$. We know that in equilibrium
(which is assumed for the sites on the backbone), the average
occupations obey Fermi-Dirac statistics \cite{amir_glass}, therefore
we know that site $A$ is negatively charged and site $B$ is
positively charged. Therefore there is a net dipole moment with a
component in the positive $\hat{x}$ direction, see Fig.
\ref{polaron_fig}. On average, it is plausible that there will
therefore be a tendency for the dipole moment of the relaxation
cluster (encircled by an eclipse in the figure) to be aligned in the
$-\hat{x}$ direction. Therefore, on average, in the relaxation
process the cluster will contribute to the electric field in the
vicinity of the sites $A$ and $B$ in the positive $\hat{x}$
direction, thus making the energy difference between them larger
still, and increasing the resistance of Eq. (\ref{R}). It is
possible that the argument does not rely explicitly on the specific
form of the interactions, as long as it is not too short ranged.
Thus, it seems like a robust and generic mechanism (albeit,
non-rigorous in the form given here) for explaining why the
conductance decreases as the system relaxes, which can be used to
explain qualitatively some of the experiments described earlier.

\subsection {Anomalous field effect and two-dip experiments - qualitative theory}
\label{anomalous_theory}

Following the theoretical discussion, we can go back to explain
qualitatively the physics behind the anomalous field effect and the
two-dip experiment.

It is clear from the argument of section (\ref{polaron}), that the
conductance decreases as the system equilibrates. Already after
single particle transitions take place, the Coulomb gap is formed,
as discussed in section (\ref{conductance}) which accounts for most
of the reduction in conductance. Nevertheless, as many particle
rearrangements take place, the system conductances continues to
decrease, which explains the long timescales involved.

What happens when we make a gate voltage scan? Let us consider a
"fast" scan, without yet mentioning compared to what. It is clear
that at a given point of the scan, the system is nearly at
equilibrium with respect to the gate voltage at which it was
equilibrated, not at the current value of the gate voltage. Thus,
the system is out-of-equilibrium with regard to the current gate
voltage, and therefore has an improved conductance. For small
changes of gate voltage, such that the system configuration is still
approximately in a local minimum (metastable state), this statement
is not true: for these, the system is nearly in equilibrium, and the
value of the conductance is nearly the lowest one possible.

This explains qualitatively why one obtains an anomalous field
effect, without explaining the characteristic scales involved, the
dependence of the scan rate, and without considering the normal
field effect which will give an additive linear dependence to the
anomalous effect.

It should be emphasized that the Coulomb gap gives intuition of how
interactions diminish the conductance of the system, but looking at
the time scales involved it is more plausible that it is formed
"immediately" at each point of the scan, and that the anomalous
field effect is due to more than single particle transitions (as mentioned earlier, Ref.
[\onlinecite{pazy}] shows that the Coulomb gap due to single
electron tunneling forms in very short times).

Let us proceed to the explanation of the two-dip experiment,
described in section (\ref{two-dip}), which is relatively
straightforward on the qualitative level once the anomalous field
effect is understood. If we make a gate voltage scan immediately
after the gate voltage is changed to its new value, clearly we are
still measuring the anomalous field effect, since the system did not
change its configuration yet. The result of a scan made after a very
long time is also clear: the system will equilibrate at the new gate
voltage, and we will measure an anomalous field effect shifted to
the new value of gate voltage. Thus, over time the initial dip must
slowly vanish, and a new one must begin to form. It turns out
experimentally that in many cases there is a symmetry between the
depth of the new dip formed and the amplitude by which the old dip
has become smaller. This can be readily understood based on the
results of section (\ref{IRM_theory}): in a two-dip protocol, the
system is initially in equilibrium, and then the gate voltage is
changed to a new value. Let us assume a quick scan is made after
time $t_w$. If we take from this scan only one value, the
conductance at the new gate voltage $V_2$, which is exactly the
position of one of the two dips, we obtain a similar dependence to
that of the IRM protocol on time during stage II of the IRM protocol
(the only difference is that here there is an additional
perturbation made, during the time of the scan itself, $t_{scan}$,
which slightly destroys the new dip formed, and has to be accounted
for \cite{grenet_thanks}). We therefore find that the new dip is
formed with an amplitude growing logarithmically in time. Let us
denote by $t_{dip}$ the time it takes to scan the width of the cusp.
Therefore measuring the conductance at the old gate voltage in the
two-dip experiment is approximately equivalent to the IRM protocol
in Stage III, with a relevant time of the order of $t_{dip}$: this
is the time the system effectively had to dig the gap at $V_1$.
Thus, the result should be $\log(1+t_w/t_{dip})$, and for $t_w \gg
t_{dip}$ the formation of the new dip and the erasure of the old one
both scale as $\log(t_w)$, which accounts for the symmetry described
above. Now we are in a position to understand the timescale involved
in the two-dip experiment, mentioned in section (\ref{two-dip}): the
requirement that the dips are of the same depth reduces to the
equation:

\be
|\log(\tau/t_{dip})|=\log(\tau/t_{dip})=|\log(\lambda_{\rm{min}}\tau)- \log(t_{scan}/t_{dip})|=-\log(\lambda_{\rm{min}}\tau t_{dip}/t_{scan}),\ee

where the term $- \log(t_{scan}/t_{dip})$ accounts for the reduction of the new dip formed due to the scan.

This finally leads to the relation:

\be \tau \sim \sqrt{\frac{t_{scan}}{\lambda_{\rm{min}}}}. \label{timescale_log}\ee

Typically, $t_{scan}$ is of the order of tens of seconds \cite{zvi:2}. Thus measuring a value of $\tau$ of the order of 1000 seconds implies an associated
timescale $1/\lambda_{\rm{min}}$ of about a day. This is possibly due to the
finite amount of time the system was let to equilibrate prior to the
experiment, and not the true cutoff of the underlying distribution.

Before continuing with the two-dip experiment, it is useful to
discuss now another way of measuring this timescale, which
experimentally leads to similar results \cite{zvi:TempProtocols}, and we shall show that it
is also consistent theoretically: in this method, one changes the
gate voltage, and then simply waits until the conductance deviation
decays to its half value from that measured \emph{after one second}.

\begin{figure}[b!]
\includegraphics[width=0.5\textwidth]{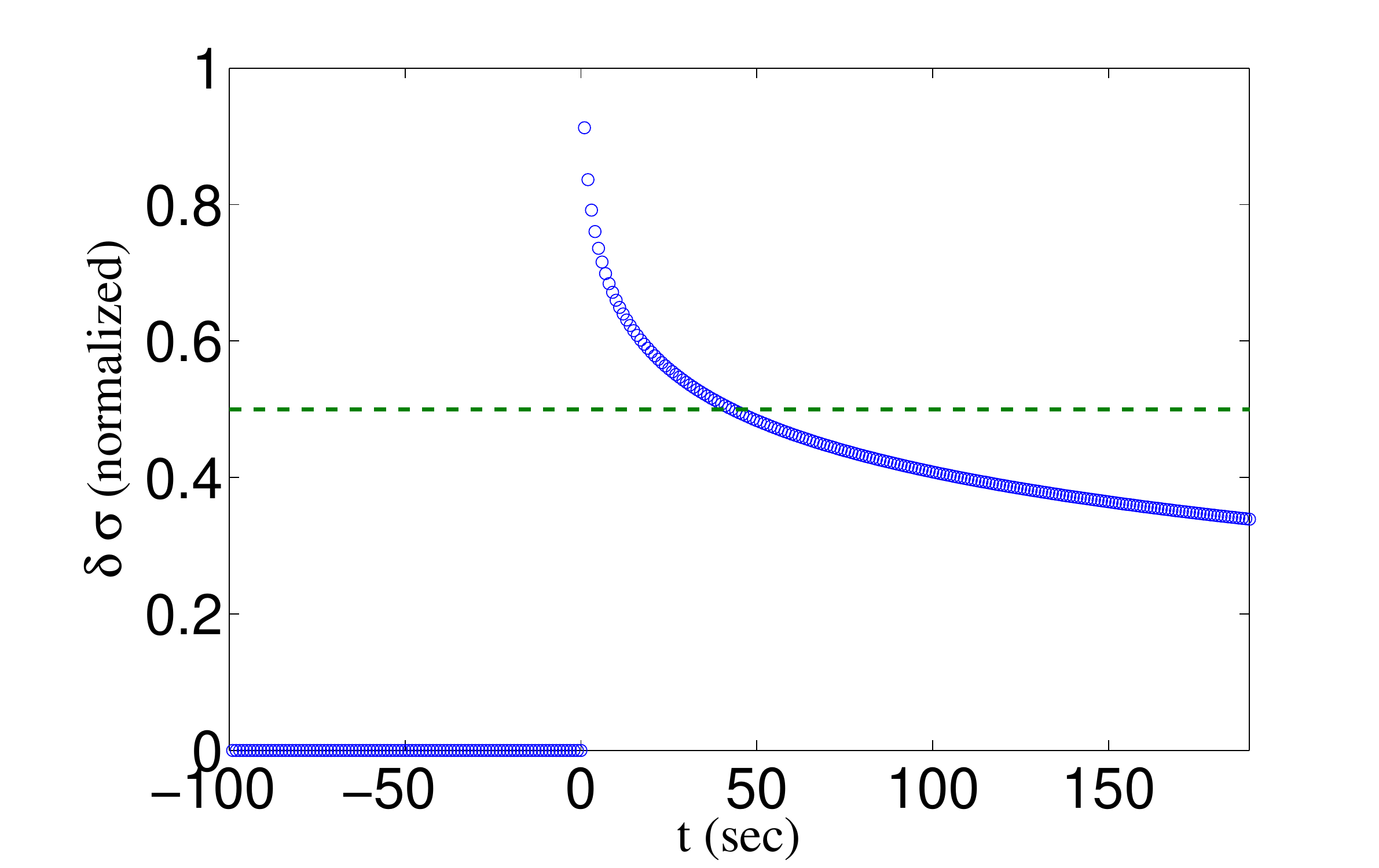}
\caption {Illustration of one of the procedures that can be used to determine the timescale of the system: the system is let to equilibrate for a long time, after which a sudden change in gate voltage is made. The signal (conductance change) is normalized to unity at $t=1$ second, and the time at which the (logarithmically relaxing) conductance reaches the value of half defines the timescale of the system. Eq. (\ref{timescale_log}) shows the physical significance of this timescale.} \label
{timescale_schematic}
\end{figure}

The procedure is demonstrated in Fig. \ref{timescale_schematic}

Here, the analysis is straightforward: according to Eq.
(\ref{log_relax}) the decay in this procedure goes as
$-\gamma_E-\log(\lambda_{\rm{min}}t)$, thus, the normalized decay is:

\be \frac{\delta \sigma(t)}{\delta \sigma(t=1
sec)}=\frac{-\gamma_E-\log(\lambda_{\rm{min}}t)}{-\gamma_E-\log(\lambda_{\rm{min}})}
.\ee

Equating this to $1/2$, we find that:

\be \tau \sim \sqrt{\frac{1 sec.}{\lambda_{\rm{min}}}}. \ee

Clearly, due to the choice of normalization after \emph{one second},
similar to the value $t_s$, the two methods qualitatively agree.

Let us return to the two-dip experiment. The theory described in
section (\ref{generalized_theory}) might hint that one can use
linear response to understand also the two-dip protocol: if the
response to a generic period sequence of square pulses can be
decomposed into a superposition of contributions arising from the
changes in gate voltage, \emph{i.e.}, $\delta \sigma (t)
 \sim \int G(t-t') \frac{dV}{dt}(t')$, where $G(t-t')$ is a function
 describing the response of the system to a step function in gate voltage, why not
 plug in a more complicated dependence of gate voltage on time,
 $V(t)$, to understand two-dip experiments quantitatively? This
 approach is incorrect, due to a fundamental reason: the dependence
 on gate voltage is not linear. For square pulses, only two values
 of gate voltage play a role, and therefore the assumption of
 linearity is unnecessary, which is why the analysis of section
 (\ref{generalized_theory}) is correct. But for the two-dip
 experiment, if we are interested in the values of the conductance
 associated with other values of gate voltage than $V_1$ or $V_2$,
 we cannot use this linear response equation.

Therefore within the framework of the $1/\lambda$ relaxation rate distribution we cannot explain quantitatively the shape of the dips, as well as other interesting
questions regarding it, e.g: What determines their dependence
temperature and disorder?

Various works have addressed these questions, which we briefly
mention now. The general physical picture presented here is similar
to that of Refs. [\onlinecite{burin, galperin}] with regard to the
mechanism leading to the decrease of the conductance and the
anomalous field effect, which is the basis for understanding
(qualitatively) the two-dip experiment. Ref. [\onlinecite{burin1}]
suggests that the slow relaxing modes are not (as was assumed above)
an intrinsic part of the electronic system, but rather, atomic
configurations which change over time and interact with the
electronic configuration. Ref. [\onlinecite{galperin}], by the same
authors, explain that the same mechanism would also be consistent
with an intrinsic picture, similar to that described here, and refer
to a specific subset of relaxation modes, with a definite structure
in space, which they refer to as "chessboard cluster". These were
introduced by the same authors in Ref. [\onlinecite{burin2}] in
order to explain the $1/f$ noise which occurs in electronic glasses,
which we discuss in section (\ref{noisesect}). Ref.
[\onlinecite{muller}] also deals with the two-dip experiment, but
focuses more on the dependence of the dip on temperature, carrier
density and disorder, and less on the timescales associated with the
formation of the phenomenon.

\subsection {Estimating the timescales} \label {timescales}

So far we have mainly discussed the functional form of the
relaxations involved, but have not demonstrated why the long time
cutoffs of the underlying distribution exceed the experimental
timescales involved (which can be hours and days). In fact, this
remains a difficult question, for which no clear cut answer exists.
While many theoretical mechanisms that can give rise to such slow
relaxations exist, it is not clear what distinguishes the slow
relaxations described in the previous section, on Anderson
insulators and granular systems, and various experiments performed
on semiconductors, for which the experimental results are different,
and no slow relaxations with full aging is observed. An experimental
clue in the direction of solving this profound problem is given by
Ref. [\onlinecite {zvi:2}], suggesting that the electron density
plays a role in determining the system relaxation time. Indeed, the electronic density in doped semiconductors below the metal-insulator transition does not reach the values of indium oxide. Recent experiments on semiconductors shows that these systems may also exhibit rich behavior and non-exponential relaxations, albeit on much shorter timescales \cite{armitage}. Measuring the density dependence of the relaxation timescales involved might shed new light on this problem. For the indium oxide samples, on the other hand, it would be good to check experimentally if for the low-density samples one can see a deviation from the clean
logarithmic relaxations, signalling that we are nearing the cutoff
of the relaxation rate distribution. An idea suggested by Ovadyhau
\cite{Ovadyahu_Tdep} emphasized the effect of Anderson orthogonality
catastrophe is slowing down the rate: indeed, experiments suggest
that the slow relaxation regime corresponds to that where more than
a single electron is present in a localization volume \cite
{zvi_extrinsic,zvi_private}. Another ingredient which may be
important for understanding the source of the slow relaxations is
the simultaneous tunneling of more than one electron
\cite{manybody1,manybody2,manybody3,manybody4,manybody5,manybody6,manybody7,efros_SCE,
pollak_manybody, galperin, amir_PR}. However, such many-particle
tunneling events can be present also in semiconductors: one has to
show that their significance depends on a parameter which is
different in the two systems - a natural candidate is the ratio of
the localization length to the average nearest-neighbor distance.
This criterion is equivalent to the above mention criterion of
having many particles in a localization. Showing this systematically
by analyzing the statistics of the many-particle tunneling processes
would be a significant step.

\section {Future prospects}

In this review we discussed the current experimental and theoretical
understanding of the memory and aging associated with electron
glasses. Throughout, a number of open questions, mostly theoretical,
were mentioned. Perhaps the most fundamental regards the timescales
associated with the relaxations. What is the role of many particle
transitions in determining it? How many electrons tunnel
simultaneously? Does the Anderson Orthogonality Catastrophe (or a
Franck–Condon type effect) play a substantial role?

Another deep question regards the basic ingredients needed to
observe the form of aging and slow relaxations discussed in this
review. Experiments performed on spin glasses, for example, show
more complex aging behavior \cite{spinglass_aging4,vincent}. It is
in our opinion a worthwhile question to understand the similarities
and differences between these two related systems. Structural
glasses also show similar behavior \cite{structural_ludwig}, which
should be further explored.

More technical questions, to which at present we do not have a full
answer, regard the quantitative understanding of the two-dip
experiment (section (\ref{two-dip}). Also, the TRM protocol
described in section (\ref{TRM_sect}) still has to be explored
further.

Altogether, it seems that electron glasses still pose many more
questions and challenges, and prove a useful platform for
investigating glassy behavior.

\section {Acknowledgements}

It is our pleasure to thank J. Delahaye, T. Grenet, M. M\"uller, Z.
Ovadyahu, M. Palassini and M. Pollak for illuminating discussions
and for useful comments regarding the manuscript. We also thank J.
Delahaye, T. Grenet and Z. Ovadyahu for providing us with their
experimental data. This work was supported by a BMBF DIP grant as
well as by ISF and BSF grants and the Center of Excellence Program.

\end{document}